\documentclass[utf8]{frontiersSCNS} 

\usepackage{aas_macros}
\usepackage{natbib}

\usepackage{url,hyperref,lineno,microtype,subcaption}
\usepackage[onehalfspacing]{setspace}

\def\msun{\hbox{${\rm M}_{\odot}$}}
\def\rsun{\hbox{${\rm R}_{\odot}$}}

\hypersetup{
  colorlinks=true,
  citecolor=blue,
  linkcolor=blue,
  urlcolor=blue
}

\def\keyFont{\fontsize{8}{11}\helveticabold }
\def\firstAuthorLast{Teyssier \& Commer\c con} 
\def\Authors{Romain Teyssier\,$^{1,*}$ and Beno\^it Commer\c con\,$^{2}$}
\def\Address{$^{1}$Centre for Theoretical Astrophysics and Cosmology, Institute for Computational Science, University of Zurich, Switzerland \\
$^{2}$Univ Lyon, Ens de Lyon, Univ Lyon1, CNRS, Centre de Recherche Astrophysique de Lyon UMR5574, F-69007, Lyon, France }
\def\corrAuthor{Romain Teyssier}

\def\corrEmail{romain.teyssier@uzh.ch}

\begin{document}
\onecolumn
\firstpage{1}

\title[Numerical Methods in Star Formation ]{Numerical Methods for Simulating Star Formation} 

\author[\firstAuthorLast ]{\Authors} 
\address{} 
\correspondance{} 

\extraAuth{}

\maketitle

\begin{abstract}
We review the numerical techniques for ideal and non-ideal magneto-hydrodynamics (MHD) used in the context of star formation simulations.
We outline the specific challenges offered by modeling star forming environments, which are dominated by supersonic and super-Alfv\'enic 
turbulence in a radiative, self-gravitating fluid. These conditions are rather unique in physics and engineering and pose particularly severe restrictions
on the robustness and accuracy of numerical codes. One striking aspect is the formation of collapsing fluid elements leading to the formation of singularities
that represent point-like objects, namely the proto-stars. Although a few studies have attempted to resolve the formation of the first and second Larson's cores, 
resolution limitations force us to use sink particle techniques,
with sub-grid models to compute the accretion rates of mass, momentum and energy, as well as their ejection rate due to radiation
and jets from the proto-stars. 

We discuss the most popular discretisation techniques used in the community, namely smoothed particle hydrodynamics, finite difference and finite volume methods,
stressing the importance to maintain a divergence-free magnetic field. We discuss how to estimate the truncation error of a given numerical scheme, 
and its importance in setting the magnitude of the numerical diffusion. This can have a strong impact on the outcome of these MHD simulations, where both viscosity and resistivity are
implemented at the grid scale. We then present various numerical techniques to model non-ideal MHD effects, such as Ohmic and ambipolar diffusion, as well as the Hall effect.
These important physical ingredients are posing strong challenges in term of resolution and time stepping. For the latter, several strategies are discussed to overcome the limitations due to
prohibitively small time steps. 

An important aspect of star formation simulations is the radiation field. We discuss the current state-of-the-art, with a variety of techniques
offering pros and cons in different conditions. Finally, we present more advanced strategies to mitigate the adverse effect of finite numerical resolution, which are very popular in the context of 
supersonic, self-gravitating fluids, namely adaptive mesh refinement, moving meshes, Smoothed Particle Hydrodynamics and high-order methods. Advances in these three directions are likely to trigger immense progress in the future of our field. We then illustrate the different aspects of this review by presenting recent results on supersonic MHD turbulence and magnetised collapse calculations.
\tiny
 \keyFont{ \section{Keywords:} star formation, numerical techniques, MHD: ideal, MHD: non-ideal, astrophysical fluid dynamics, radiation fields, sink particles} 
\end{abstract}

\section{Introduction}

Star formation is one of the main unsolved problems in astrophysics. Although our view of this fundamental 
process, at the nexus of galaxy formation, planetary science and stellar evolution, has considerably changed
over the past decades, thanks to new observations and theoretical progress, many dark corners remain to be explored.
One of the reasons why the true origin of stars still eludes us is the highly non-linear nature of the governing equations,
describing self-gravitating, compressible, magnetized dust and gas fluids interacting with radiation. In this review, we
present these main governing equations, focusing on ideal magneto-hydrodynamics, radiation hydrodynamics, non-ideal effects
and sink particle techniques. We describe the most popular discretisation schemes, focusing on possible sources
of errors and their consequences on clouding our conclusions. We finally give a short description of the landscape 
in term of star formation simulations, with different set-up strategies addressing particular scales in this
fundamentally multi-scale problem.

Star formation is believed to be the consequence of the collapse of mildly supersonic or transonic to subsonic molecular cores 
emerging out of supersonic turbulent flows in the interstellar medium \citep{2004RvMP...76..125M,2007ARA&A..45..565M}. 
The source of turbulence is probably to be found on large scales, as a consequence
of large galactic shearing or colliding flows, but also on small scales, because of various sources of stellar feedback 
\citep{1989ApJ...345..782M,2017IAUS..322..123F}. 
In this context, gravitational or thermal instabilities lead to the formation
of dense gas clumps that undergo a more or less violent gravitational collapse, leading to the formation of a proto-star
surrounded by a proto-stellar disk. Describing these processes using only simple analytical methods is almost impossible. Moreover,
the traditional engineering methods in Computational Fluid Dynamics (CFD) are usually not robust enough to sustain
highly supersonic and super-Alfv\'enic flows. Self-gravity, magnetic fields and radiation fields, taken together, define
a very unique system of equations that has no equivalent in the industry. This has led astrophysicists to develop their own 
methods, largely inspired by traditional methods designed in the applied mathematics community, but significantly adapted 
to the specific problem at hand. In this context, Smoothed Particle Hydrodynamics (SPH) and Adaptive Mesh Refinement (AMR) techniques
turned out to be particularly suitable to the problem of star formation. Both techniques have their own pros and cons, and comparing 
the two allows us to assess the robustness of our numerical results. New techniques have also been developed, that are specific to
star formation, such as sink particles, a commonly adopted recipe to replace an otherwise collapsing fluid element
by a collision-less particle, saving computational times and increasing the realism of the simulation. 

In this review, we pay attention to the description of the equations, without necessarily discussing their physical foundations, 
such as the ideal MHD limit or the non-ideal diffusion processes. We describe the various numerical techniques, from low-order
schemes to modern high-order methods, as well as from non-zero divergence schemes to exact divergence-free methods, etc. 
We refer to the corresponding literature, including all references that are relevant to the historical description of the discipline
and that give a fair snapshot of the present state of the field. We apologize in advance for not having included all possible 
references on the topic.

\section{Ideal MHD: numerical techniques}
\label{imhd}

Developing new computational methods for solving the ideal MHD equations has generated a lot of interest 
within the applied mathematics and computational physics communities. Quite naturally, because of their
success in solving the Euler equations, grid-based methods with flux upwinding, also known as Godunov's method, 
were applied to the MHD equations in the framework of finite-volume discretisation. In parallel, Smoothed Particle Hydrodynamics (SPH) 
generated a lot of interest in the astrophysics community because of its strict Galilean invariance. Both methods,
however, quickly ran into difficulties trying to maintain the divergence-free property of the MHD equations.

\subsection{The ideal MHD equations}

Before we review the various improvements of MHD numerical solvers over the past decades, we briefly recall 
the ideal MHD equations, shown here in conservative form. We have first the mass conservation equation,
\begin{equation}
\frac{\partial \rho}{\partial t} + \nabla \cdot \left( \rho {\bf v}\right) = 0,
\end{equation}
where $\rho$ is the mass density and ${\bf v}$ is the fluid velocity vector. We also have the momentum
conservation equation
\begin{equation}
\frac{\partial }{\partial t} \left( \rho {\bf v}\right)  + \nabla \cdot \left( \rho {\bf v} \otimes {\bf v} + P_{\rm tot} {\mathbb I} 
- \frac{1}{4\pi}{\bf B}\otimes{\bf B}\right) = \rho {\bf g},
\end{equation}
where ${\bf B}$ is the magnetic field vector and $P_{\rm tot}$ is the total pressure, defined as the sum of the thermal pressure
and the magnetic pressure
\begin{equation}
P_{\rm tot} = P + \frac{1}{8\pi}B^2.
\end{equation}
Note that we work here in {\it cgs} units, hence the presence of the $4\pi$ term in these equations.
We have also included the gravitational acceleration vector ${\bf g}$ as a source term on the right-hand side of the momentum conservation equation.
Finally, we have the total energy conservation equation
\begin{equation}
\frac{\partial E_{\rm tot}}{\partial t} + \nabla \cdot \left( \left( E_{\rm tot}+P_{\rm tot}\right) {\bf v} - \frac{1}{4\pi} {\bf B}({\bf B}\cdot{\bf v}) \right) = \rho {\bf g} \cdot {\bf v},
\end{equation}
where the total fluid energy is the sum of the kinetic energy, the internal energy and magnetic energy
\begin{equation}
E_{\rm tot} = \frac{1}{2}\rho v^2 + e + \frac{1}{8\pi}B^2.
\end{equation}
In order to close the system, we need the fluid equation of state, usually given be the ideal gas equation of state
\begin{equation}
P=(\gamma-1)e,
\end{equation}
and the induction equation for the evolution of the magnetic field
\begin{equation}
\frac{\partial {\bf B}}{\partial t } = - \nabla \times {\bf E}{\rm ~~~with~~~}{\bf E}= - {\bf v} \times {\bf B},
\label{eq:induction}
\end{equation}
where we have introduced the electric field ${\bf E}$ (also known as the electromotive force, emf). 
We need to add to these equations the most important property of the magnetic field, namely
\begin{equation}
{\nabla} \cdot {\bf B}=0,
\end{equation}
also known as the solenoidal constraint or the divergence-free condition.
Since we consider now in this section ideal MHD conditions, we have no dissipation terms in the induction equation, 
as well as in the fluid equations.

\subsection{Some important mathematical properties}

The fluid equations appear to be in conservative form, which naturally calls for a finite volume representation,
which will ensure exact conservation of mass, momentum and total energy, by construction, owing to the divergence theorem
\begin{equation}
\int_V \nabla \cdot {\bf F} {\rm d}V = \int_S {\bf F}\cdot {\bf n}{\rm d}S,
\end{equation}
where the vector ${\bf F}$ can be the flux of mass, momentum or energy.
Designing a numerical scheme boils down to computing the flux through the surface of the volume elements. 
On the other hand, the induction equation naturally calls for a finite surface representation,
owing to the Stoke's theorem
\begin{equation}
\frac{\partial}{\partial t} \int_S {\bf B}\cdot {\bf n}{\rm d}S = \int_S \nabla \times {\bf E} \cdot {\bf n}{\rm d}S = \int_L {\bf E} \cdot {\bf {\rm d}l}.
\end{equation}
Similarly, the divergence-free condition, written in integral form as
\begin{equation}
\int_V \nabla \cdot {\bf B}{\rm d}V = \int_S {\bf B}\cdot{\bf n}{\rm d}S,
\end{equation}
also calls for defining the magnetic field as a surface averaged quantity.
This naturalness argument, together with the fact that the divergence-free condition can be maintained
to machine precision accuracy, has led to the design of the {\it constrained transport} scheme \citep{1988ApJ...332..659E}.
This very popular method for grid-based techniques comes however with a price: the flow variables are not all co-located at cell
centers, like volume-weighted fluid variables, but also at face centers, for the magnetic field components, and at edge centers
for the electric field components. 

A very important mathematical property of the magnetic field that emerges from the divergence-free condition is that the normal component 
of the field should be continuous across cell faces. The $x$-component of the field, for example, can be discontinuous in the $y$ and $z$-directions,
but has to vary smoothly in the $x$-direction. This property also naturally holds in the constrained transport method. It can be written as
\begin{equation}
\int_S  \left[ {\bf B}\cdot {\bf n}\right]{\rm d}S = 0,
\end{equation}
where the operator $[A]=A^+-A^-$ denotes the jump of a quantity $A$ across the surface element.

\subsection{Preserving the divergence-free condition}

Historically, one of the first ideal MHD, general purpose codes, is ZEUS-2D 
developed specifically for astrophysics by Stone and Norman \citep{1992ApJS...80..791S}.
It is a finite-difference code using constrained transport and artificial viscosity to handle shocks, The continuity and divergence-free conditions are therefore 
satisfied by construction, but since artificial viscosity is used to handle shocks,  instabilities can occur in fast rarefaction waves \citep{2002ApJ...577L.123F}.
 
The other popular strategy for grid-based methods is to maintain all MHD variables as volume-averaged quantities, allowing for discontinuities
across all cell faces. In the late 90s, a series of papers presented finite-volume
MHD codes with proper upwinding of numerical fluxes using Riemann solvers \citep{Dai:1994iv,Ryu:1995bx,1996ApL&C..34..245T,1998ApJS..116..133B,2009JCoPh.228..952L}.
These methods are very powerful, because they are strictly conservative and because they satisfy the so-called {\it entropy condition}, meaning the entropy
can only increase, a key property to maintain the stability of the numerical solution.
These finite-volume codes all considered the magnetic field as a volume-averaged, piecewise-constant, cell-centered quantity, which violates the continuity
condition of the normal component of the field. The resulting schemes are therefore not necessarily divergence-free anymore. 
In order to maintain the divergence as small as possible, an additional step is required that modifies the magnetic field components and decreases or nullifies the divergence: this operation is called {\it divergence cleaning}. 
In a  seminal paper, \cite{2000JCoPh.161..605T} has compared various schemes and showed that they all passed with mixed success a variety of MHD test
problems. We will now review these grid-based methods that are using divergence cleaning.

The first method we discuss here is the {\it projection scheme}, introduced by \cite{1980JCoPh..35..426B}. The idea is to measure 
the spurious divergence after the main update of the MHD variables, and solve for a Poisson equation defined as
\begin{equation}
\Delta \phi = \nabla \cdot {\bf B}^*.
\end{equation}
where $\phi$ is a scalar potential. The new magnetic field is then corrected as
\begin{equation}
{\bf B}^{n+1} = {\bf B}^* - \nabla \phi,
\end{equation}
so that by construction the divergence is now zero. This method works very well in many applications.
It suffers however from two main issues: first, it is quite expensive as it requires to solve for a Poisson equation. 
Another consequence is that the correction is non-local: very localized divergence errors can be instantaneously transported
across the grid to enforce the solenoidality condition. Second, the correction process modifies the magnetic field,
without modifying the total energy. As a consequence, the resulting temperature can be modified, and the entropy condition might be violated.
An easy fix is to remove the magnetic energy before the correction step and add the new magnetic energy after the correction, 
but the resulting scheme is not conservative anymore.

The second popular method is the so-called 8-waves formulation or Powell's method \citep{1999JCoPh.154..284P}. The idea is to write
more general ideal MHD equations {\it allowing for the presence of magnetic monopoles}. This results in the following non-conservative form
\begin{align}
\frac{\partial }{\partial t} \left( \rho {\bf v}\right)  + \nabla \cdot \left( \rho {\bf v} \otimes {\bf v} + P_{\rm tot} {\mathbb I} 
- \frac{1}{4\pi}{\bf B}\otimes{\bf B}\right) &= - \left( \nabla \cdot {\bf B}\right) {\bf B} + \rho {\bf g}, \\
\frac{\partial E_{\rm tot}}{\partial t} + \nabla \cdot \left( \left( E_{\rm tot}+P_{\rm tot}\right) {\bf v} - \frac{1}{4\pi} {\bf B}({\bf B}\cdot{\bf v}) \right) &= 
- \left( \nabla \cdot {\bf B}\right) {\bf B} \cdot {\bf v} + \rho {\bf g} \cdot {\bf v},\\
\frac{\partial {\bf B}}{\partial t } - \nabla \times \left( {\bf v} \times {\bf B}\right)  &= - \left( \nabla \cdot {\bf B}\right) {\bf v}.
\end{align}
This method proved very useful and robust for many applications. It is still widely used in astrophysics nowadays.
The success of the method originates from the property that the spurious $\nabla \cdot {\bf B}$ is advected away with the flow,
using the so-called eighth MHD wave, so that divergence errors do not accumulate. There are however two problems: First, in stagnation points\footnote{Stagnation points are regions where the flow is brought to rest in the frame of the system under study. More generally, even if the velocity does not vanish, stagnation points are regions in the flow where the streamlines are converging and the flow becomes compressive.}, 
the divergence errors will accumulate because the flow trajectories are converging. Second, the scheme is not strictly conservative. Shock waves lead to solutions that do not converge to the correct conservation laws anymore. 

A very elegant solution to the first problem was proposed in \cite{2002JCoPh.175..645D} using the so-called hyperbolic-parabolic divergence
cleaning technique, also known as a Generalized Lagrange Multiplier (GLM) formulation of the ideal MHD equations, in short, {\it Dedner's scheme}. 
The idea is to add a ninth wave to the previous Powell's modified MHD equations, introducing the cleaning field $\psi$ that satisfies
\begin{equation}
\frac{\partial \psi}{\partial t} + {\bf v}\cdot \nabla \psi + c_h^2 \nabla \cdot {\bf B} = - \frac{\psi}{\tau},
\end{equation}
and is used as a source term in the induction equation.
\begin{equation}
\frac{\partial {\bf B}}{\partial t } - \nabla \times \left( {\bf v} \times {\bf B}\right)  = - \nabla \psi.
\end{equation}
In the first equation, $c_h$ is the divergence wave speed and $\tau$ is the divergence damping time scale.  
The latter is chosen equal to (or larger than) the fast magnetosonic wave speed, while the former is usually equal to
(or larger than) the fast magnetosonic cell crossing time.
At first sight, this new nine waves scheme can be seen as a combination of advection and damping of divergence errors, 
thus a clever combination of the projection method and the Powell's scheme. 
As shown recently by \cite{2018JCoPh.364..420D}, it is in fact much more than that: 
this new divergence field $\psi$ allows to restore the entropy condition, 
as this field can be interpreted as a divergence cleaning energy which stores temporarily the magnetic energy lost during the damping step.
It is also conservative in a general sense, but still violates the Rankine-Hugoniot shock relations locally. 

In parallel, accurate and stable MHD solvers have been the topics of many studies in the SPH framework.
Probably because of its strict Galilean invariance, early SPH MHD solvers were quite oscillatory \citep[see e.g.][]{1999A&A...348..351D}. 
Truncation errors associated to the non-zero divergence cannot be damped by numerical diffusion as they are advected away,
as in the case of grid-based codes. Many regularization techniques have been proposed to provide stable SPH solvers for the ideal MHD equations:
using the vector potential \citep[e.g.][]{Kotarba:2009eb} or using artificial diffusivity \citep{Price:2005kx,dolag:09}.
Interestingly, the work of \cite{Price:2005kx} was based on exploring various divergence cleaning methods used in grid-based codes for SPH.
The authors concluded that this works in most cases, but in difficult cases, such as supersonic turbulent flows, this can lead to spurious
effects, such as the violation of the entropy condition or wrong shock relations. More worrysome, these authors noticed that divergence
cleaning can lead to an increase of the local magnetic energy, a price to pay to redistribute the truncation errors in the divergence.
This results in {\it spurious dynamo effects.} Recently, however, \cite{Tricco:2012ju} revisited the Dedner's scheme for SPH and found
a formulation that guarantees that divergence cleaning leads to a decrease of the magnetic energy and an increase of the entropy, 
in the spirit of \cite{2018JCoPh.364..420D} for grid-based codes.

\begin{figure}[h!]
	\begin{center}
		\includegraphics[width=17cm]{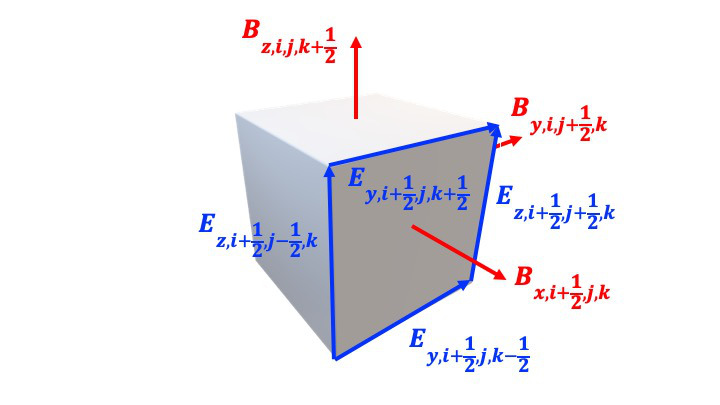}
	\end{center}
	\caption{Schematic showing the geometry of a Cartesian cell in the Constrained Transport approach. The finite volume cell is shown as a grey cube.
It is labelled $i,j,k$. The magnetic field components are defined perpendicular to the faces of the cube. The are shown in red, only in the rightermost face in all three directions.
The electric field components are defined on the cell edges. They are shown in blue and only in one face for sake of simplicity.}
        \label{fig:cube3d}
\end{figure}

In light of these rather complex developments of divergence cleaning methods, the constrained transport (CT) approach we have introduced already
seems quite appealing. Note however that this approach requires to have well defined cell-interfaces, which is not necessarily the case for 
SPH or the recently developed moving-mesh AREPO code \citep{2010MNRAS.401..791S}. The CT scheme, introduced for astrophysical fluid flows 
by \cite{1988ApJ...332..659E} and used in the ZEUS-2D code \citep{1992ApJS...80..791S}, features the nice property that the divergence of the magnetic
field, defined in integral form as the net magnetic flux across the 6 faces of each cell, will always remain constant, So if it is initially zero,
it will remain equal to zero within numerical round-off errors. This however requires to define the electric field on the cell edges,
and this is the main difficulty of this approach (see Figure~\ref{fig:cube3d}). Indeed, in the Godunov methodology, fluxes, defined at cell faces, are computed as the solution 
to the Riemann problem of the two neighboring cells meeting at their common cell interface. For the electric field, defined at cell edges,
4 neighboring cells meet and in order to properly upwind the electric field, we need to solve a {\it two-dimensional Riemann problem}, 
a rather daunting task \citep[see][for a discussion]{2006JCoPh.218...44T}.
Several solutions have been found to design 2D MHD Riemann solvers \citep{2004JCoPh.195...17L,2014JCoPh.261..172B}, and this has been the main characteristic
of several simulations codes with proper upwinding of both fluxes and electric fields, using the CT scheme within the Godunov methodology 
\citep{2004JCoPh.195...17L,2006A&A...457..371F,Stone:2008vp,2012ApJ...745..139L,2014MNRAS.442...43M}.

\subsection{Minimizing numerical errors: higher order versus mesh refinement}

When performing self-gravitating magnetized fluid simulations, it is of primary importance to quantify and understand numerical errors.
These errors are often called {\it truncation errors} because they arise as leading order terms in the Taylor expansion of the adopted 
discrete numerical method. Usually, this Taylor expansion leads to the so-called {\it modified equation}, which encodes how the original 
ideal MHD equations have been transformed into a new set of equations with additional terms coming from the truncation errors.
For first-order schemes, these terms are identical to a diffusion process, and are therefore called {\it numerical viscosity}
or {\it numerical diffusivity.} In SPH, quite often, these terms are added explicitly to the equations as {\it artificial viscosity}
and {\it artificial diffusivity}, while for grid-based Godunov solvers, these terms are only implicitly present, through this Taylor
expansion of the discrete scheme. In both cases, however, these diffusion processes control how shock heating and magnetic reconnection
proceed in the solution. They play a fundamental role in preserving the entropy condition, and in regulating the flow close to the grid scale.
Unfortunately, many complex MHD processes, such as the small scale dynamo \citep{2005PhR...417....1B} 
or the magneto-rotational instability \citep{1991ApJ...376..214B} depend crucially on the 
so-called Prandtl number \citep{2007AA476.1123F,2014ApJ...797L..19F}, which is the ratio of the real magnetic diffusivity to the viscosity. In most case, this ratio is always close to 1
if one uses the numerical Prandtl number \citep{2007AA476.1123F,2011PhRvL.107k4504F}, 
while in nature, it can vary widely. It is therefore crucial to adopt in some cases explicit viscosity 
and diffusivity coefficients and model the flow including these non-ideal processes \citep{2007AA476.1123F,2016JPlPh..82f5301F}. 

As explained in the next section, in order to model these non-ideal effects, it is crucial to control and minimize the numerical
diffusion as much as possible. There are two possible strategies to achieve this: refine the grid or increase the order of accuracy of the method.
The first approach leads to the so-called Adaptive Mesh Refinement method, a very popular and successful technique in the context of
star and galaxy formation. Popular AMR codes are available to the star formation community, such as RAMSES \citep{2002A&A...385..337T,2006A&A...457..371F},
ORION \citep{2007ApJ...667..626K,Klein:1999hl}, ENZO \citep{2014ApJS..211...19B} or FLASH \citep{2000ApJS..131..273F}. 
Other codes, that used to be only unigrid, now propose an adapted grid or an adaptive grid extension, such as PLUTO \citep{2007ApJS..170..228M} and ATHENA \citep{Stone:2008vp}. 
In all these codes, cells are adaptively refined according to various refinement criteria. In the context of star formation, the most popular approach 
is to always resolve the local Jeans length with 4 cells or more, the so-called Truelove criterion \citep{1997ApJ...489L.179T}
\begin{equation}
{\rm If~}\lambda_{\rm J}= \frac{c_s}{\sqrt{4\pi G \rho}} < 4 \Delta x_{\ell}~~~{\rm then~refine~to~level~}{\ell +1}.
\end{equation}
This corresponds to a level-dependent density threshold that triggers new refinements. SPH methods are Lagrangian in nature,
so they cannot refine as much as grid-based codes. Usually, much more SPH particles are needed in the initial conditions to reach a certain 
target Jeans mass, corresponding to the maximum resolution level of the corresponding AMR simulation. Particle splitting 
is an interesting alternative to classical SPH but is still under development \citep{2002MNRAS.330..129K,2015MNRAS.451.3955C}, 
the difficulty being to handle sharp transitions in particle mass and its interaction with the smoothing kernel.
Note that if similar resolution requirements are
met, AMR and SPH methods largely agree on the quantitative predictions on how the collapse proceeds \citep{commercon:08}.
For magnetized flows, the Jeans length-based refinement strategy has to be more conservative, of the order of 30 cells per Jeans 
length, in order to capture properly the magnetic field amplification in collapsing cores \citep{2010ApJ...721L.134S,2011ApJ...731...62F,2012ApJ...745..154T}.

A difficulty arises when one uses AMR for Constrained Transport. In this case, it is mandatory to be able to interpolate the
magnetic field from the coarser level to the finer level and still satisfy the divergence-free condition. For this, 
divergence preserving interpolation schemes have been developed \citep{2001JCoPh.174..614B,2002JCoPh.180..736T} and play an important
role in the viability of the CT approach in the context of AMR.

To reduce truncation errors, the other option is to use higher order schemes. The solution 
inside each cell is not piecewise constant as in the traditional first-order Godunov method,
but it is reconstructed using high-order polynomials of degree $p$ as
\begin{equation}
\rho(x)=\sum_{i=0,p} \alpha_i \psi_i(x).
\end{equation}
The $\psi_i(x)$ are usually an orthonormal basis of polynomials of degree at most $p$.
The coefficients $\alpha_i$ are computed using two different philosophies. The first option, the WENO approach
computes the coefficients $\alpha_i$ at each time step using neighboring cells. The higher the polynomial degree,
the more neighbors must be included in the stencil of the method \citep{1999JCoPh.150..561J,2014JCoPh.275..484B}. 
The second option, the Discontinuous Galerkin (DG) approach, considers that the $\alpha_i$ are new variables, defined for each MHD 
variables, and updated using a projection of the ideal MHD equation onto the polynomial basis \citep{Li:2012cl,Klingenberg:2017ck,2014MNRAS.437..397M,Guillet:2018wr}. 
In the WENO case, one needs to access neighbouring but possibly distant cells to compute the $\alpha_i$, while in the DG case,
one needs to store permanently the $\alpha_i$ and solve for their evolution at each time step.
Note that other high-order methods are also being developed that do not strictly correspond to neither WENO nor DG \citep{2018JCoPh.375.1365F}.

In the context of high-order methods, the divergence free condition is particularly challenging. One can either 
implement high-order version of one of the above mentioned divergence cleaning techniques \citep{2016JCoPh.317..223D}, or one can try to preserve 
the divergence-free constraint within the method. The second approach could be seen as a generalisation of the 
CT scheme to higher order. The key ingredients are: 1- the use of a divergence-free polynomial basis for the 
magnetic field \citep{Li:2005bf,Guillet:2018wr}, 2- enforcing the continuity of the normal field across cells boundaries \citep{Li:2012cl}. Although some very promising 
solutions have been found recently, this is still a very active field of research.
Interestingly, the traditional CT scheme with face-centred magnetic field variables can be re-interpreted as a DG scheme
where each magnetic field component is piecewise-linear and continuous in the normal direction and piecewise constant in the transverse
direction, so that two cell-centred coefficients are required (instead of one) for each field component.

\section{Non-ideal MHD: numerical techniques}
\label{Sec:nimhd}
\subsection{Equations and basic concepts}

Star formation takes place in molecular clouds, which are made of a mixture of dust and gas, implying dust and gas collisions, and both constituents are far from being fully ionised. In the previous section, we presented the work done in the ideal MHD framework, which do not appear to be well suited to the ionisation state  in collapsing cores, in particular at the onset of disk formation. Recent works have emphasized the imperfect coupling of the dust and gas mixture with the magnetic fields at the transition between the envelop and the disk in collapsing cores (see the review by \cite{wurster:18b} for the work done in the context of protostellar disk formation) and a lot of effort has been devoted over the past ten years to include the so called non-ideal effects: the ambipolar diffusion, the Ohmic diffusion, and the Hall effect. 
The ambipolar diffusion is the common name to describe the interaction between neutrals and charged particles. It can be seen as a friction term, it enables the neutral field to respond to the magnetic forces, via collisions with charged particles. The Ohmic diffusion results from the collision of electrons with the neutrals. Last, the Hall effect is due to the drift between the positively and negatively charged species. As shown in \cite{marchand:16} and \cite{tsukamoto:15b}, {\it{all}} these three terms can be dominant over the others at different scales within the envelop of collapsing dense cores. For a classical dust size distribution, \cite{mathis:77} and \cite{marchand:16} find that ampipolar diffusion and the Hall effect dominate at densities $<10^{12}$~cm$^{-3}$ and that Ohmic diffusion is the stronger resistive effect at higher densities. 

The exact scale and density at which these resistive effects become dominant over the other dynamical processes depend on the chemistry, the ionisation intensity and the dust grain size distribution \citep{zhao:16,dzyurkevich:17,zhao:18,zhao:18b}. \cite{hennebelle:16} have shown that ambipolar diffusion regulates the flow dynamics over the other dynamical processes (induction, rotation and free-fall) at scale of a few 10s~AU, which sets the initial size of protostellar disks. In addition, \cite{masson:16} have shown that scales of a few 10s AU exhibit  magnetic Reynolds numbers less than unity (see also \cite{tomida:13,tomida:15,vaytet:18} for studies including the Ohmic diffusion). Recently, \cite{koga:19} performed a similar analysis considering the Hall effect only and found coherent results as for ambipolar diffusion.

Before we describe the numerical implementation for the three aforementioned resistive effects, let us recall the main equations and define the necessary quantities. Ideally, the system should account for the different behaviour of the neutrals, as well as negatively and positively charged particles. Among them the different constituents that participate to the ionisation balance, we have neutral (molecule, atoms, dust grains), ions (molecular and atomic), electrons, and charged dust grains. The latter can be positively or negatively charged, and can hold multiple charges \citep{draine:87}. 
Most current works follow a one-fluid approximation to describe the evolution of this complex mixture, where the Ohm's law accounts for the non-ideal effects with some assumptions. In the following, we focus on the work done in the single and two-fluid approximations. We review briefly the work done towards a multi-fluid treatment of the charged and neutral particles in Sec.~\ref{sec:multifluid}. 

In the low ionisation limit, each kind of charged particle is scarce, so that the pressure terms, the gravity, and the inertia term can be neglected in their momentum evolution equation  in comparison to the Lorentz force and the frictional force exerted by the neutrals. The generalised Ohm's law reads
\begin{equation}
{\bf E_{NIMHD}} = {\bf v}\times {\bf B} -\eta_\Omega {\bf J}  - \frac{\eta_\mathrm{H}}{|| {\bf B}||}{\bf J}\times{\bf B} + \frac{\eta_\mathrm{AD}}{|| {\bf B}||^2}{\bf J}\times{\bf B}\times{\bf B},
\label{emf_nimhd}
\end{equation}
where ${\bf J}=\nabla \times {\bf B}$ is the current, $\eta_\Omega$, $\eta_\mathrm{H}$, and $\eta_\mathrm{AD}$ are the Ohmic, Hall, and ambipolar resistivities, respectively (in units of cm$^2$~s$^{-1}$), and ${\bf v}$ is the velocity of the neutrals. The notation $||{\bf B}||$ represents the norm of the magnetic field vector ${\bf B}$. The electric field is then replaced in Eq.~(\ref{eq:induction}). It is worth noticing that all the resistive terms lead to parabolic partial differential equations to discretise. We note that only the Ohmic term can lead to reconnection here. It can be rewritten as a Laplacian operator $\eta_\Omega\Delta {\bf J}$, whereas the two others cannot. The ambipolar diffusion and the Hall effect can be assimilated to drift velocities of the magnetic fields with respect to the neutral speed ${\bf v}$. The total energy equation has to be modified accordingly, to account for the heating resulting from the friction terms, i.e., the ambipolar diffusion and the Ohmic diffusion
\begin{equation}
\frac{\partial E_{\rm tot}}{\partial t} + \nabla \cdot \left( \left( E_{\rm tot}+P_{\rm tot}\right) {\bf v} - \frac{1}{4\pi} {\bf B}({\bf B}\cdot{\bf v}) \right) = \rho {\bf g} \cdot {\bf v} + \eta_\Omega||{\bf J} ||^2 + \eta_\mathrm{AD}\frac{||{\bf J}\times {\bf B}||^2}{||{\bf B}||^2}.
\label{eq:etot_nimhd}
\end{equation}

We refer readers to \cite{shu:87}, \cite{nakano:02} and \cite{balbus:09} for more details on the derivation of the generalised Ohm's law in the one-fluid approximation. In the context of star formation, where a huge dynamical range has to be considered, the estimate of the resistivities is challenging. It requires to know the ionisation state of the gas and dust mixture. The coupled chemistry of gas and dust has to be considered, which involves a huge variety of physical processes:  gas phase and gas-dust chemistry, cosmic ray ionisation, UV radiation, (in-)elastic collisions,  grain growth. 
This domain remains the subject of intense research works that we do not detail in this review. We refer readers to Sec. 4.0.1 in the review by \cite{wurster:18b} for a scan of the works that tackle the resistivity calculations.

Ambipolar diffusion and Ohmic diffusion were the first to be considered in star formation applications. Indeed, these two terms do not introduce new MHD waves in the system (they do not change the spectral properties of the hyberpoblic system), so that they do not require heavy modifications of the MHD solver. The induction equation integration is most of the time splitted in two steps: first the ideal MHD and second the resistive terms. If an energy equation is considered, the heating terms due to resistive effect in Eq.~\ref{eq:etot_nimhd} are integrated as source terms in most numerical implementations.

\subsection{Ohmic diffusion}

The Ohmic diffusion is perhaps the simplest resistive term to introduce because of its diffusive nature, similar to the introduction of artificial viscosity and artificial resistivity to prevent any numerical artefact due to numerical diffusion.   Assuming a constant resistivity, the Ohmic term can be rewritten as a Laplacian (which preserves the solenoidal constraint), leading to the following stability condition
\begin{equation}
\Delta t_\Omega = \frac{\Delta x^2} {\eta_\Omega}.
\end{equation}
It corresponds to the stability condition required in numerical schemes integrating parabolic equations such as the heat equation with an explicit scheme.

Numerous implementations for the Ohmic diffusion have been developed in the past ten years, in all different kind of codes.  \cite{masson:12} present a fully explicit implementation in the AMR code RAMSES, where they update the emf to account for the resistive terms and then use the CT scheme as for ideal MHD. Because of the restrictive stability condition for the resistive term which can be smaller that the MHD one, some authors considered schemes that enable to  relax the time step constraint. For SPH, \cite{bonafed:11} implemented Ohmic diffusion with a constant resistivity in GADGET following the method used for artificial dissipation \citep[e.g.,][]{Price:2005kx,dolag:09}. \cite{tomida:13} implemented an explicit scheme in a nested-grid code following the implementation of \cite{matsumoto:11} in the AMR code SFUMATO. They adopt the Super-Time-Stepping (STS) algorithm proposed by \cite{alexiades:96}  to accelerate the integration of the diffusion term and to relax the stringent explicit CFL condition. 

The STS consists in a discretisation as a classical explicit scheme, which does not require to satisfy the stability in every step but after a series of integration cycles which are made of $N_\mathrm{STS}$ sub-timesteps $\Delta t_\mathrm{STS}$ which are based on Chebyshev polynomials to guarantee the stability conditions over the super-timestep $\Delta t_\mathrm{tot}=\sum_1^{N_\mathrm{STS}}\Delta t_\mathrm{STS}$. 
We also note the work of \cite{2007ApJS..170..228M,osullivan:06} where Ohmic diffusion is also integrated using the STS technique. Similarly, \cite{tsukamoto:13} proposes a SPH discretisation of Ohmic dissipation using STS. \cite{wurster:16} and \cite{price:17} (in the PHANTOM code) use similar implementations on different SPH codes. Last, \cite{marinacci:18} propose two implementations in AREPO: one using Powell divergence cleaning and another using CT. Both implementations can be integrated using an explicit or implicit scheme. Almost all the implementations mentioned so far for the Ohm diffusion are combined with ambipolar diffusion (see below).

Apart from resistive MHD, the STS technique is becoming increasingly popular in the computational astrophysics community to accelerate parabolic timestepping advancements in the context of radiation hydrodynamics with the FLD \citep{commercon:11b} and anisotropic diffusion \citep{meyer:12,vaidya:17}. 

\subsection{Ambipolar diffusion}

Ambipolar diffusion is not as straightforward to include since the associated term in the induction equation can not be rewritten as a Laplacian \citep{brandenburg:94}. For the diffuse ISM, the drift between ions and neutrals is the effect of the ambipolar diffusion, that can be written as a friction force
\begin{equation}
\bf{F}_{in}=\rho_i \rho_n \gamma_\mathrm{AD}  (\bf{v}-\bf{v}_i),
\end{equation}
where $\rho_i$ (resp. $\bf{v}_i$) and $\rho_n$ ($\bf{v}$) are the mass density (resp. velocity) of ions and neutrals, and $\gamma_\mathrm{AD}$ is the collisional coupling constant, with $\gamma_\mathrm{AD}\sim 3\times 10^{13}$~cm$^3$~s$^{-1}$~g$^{-1}$ in ISM mixture \citep{draine:83}. In the one-fluid and strong coupling limit, the inertia of ions is neglected, as well as the pressure and gravitational forces (the ionisation degree is very low). As a consequence, the Lorentz force equals the drag force so that the ion drift velocity is
\begin{equation}
\bf{v}_{d}\equiv \bf{v}_i-\bf{v} = \frac{1}{ \rho_i \rho_n \gamma_\mathrm{AD}}\left( \nabla \times \bf{B} \right) \times \bf{B}.
\end{equation}
The induction equation thus reads
\begin{equation}
\frac{\partial \bf{B}}{\partial t} =  \nabla\times [\bf{v}_i \times \bf{B}] =   \nabla\times \left[\bf{v} \times \bf{B}  +  \frac{1}{ \rho_i \rho_n \gamma_\mathrm{AD}}\left( \nabla \times \bf{B} \right) \times \bf{B} \times \bf{B}\right].
\end{equation}
From Eq. \ref{emf_nimhd}, we have $\eta_\mathrm{AD}= || \bf{B}|| ^2/\rho_i \rho_n \gamma_\mathrm{AD}$. In the dense ISM, the heavy charged particles are ions and charged dust grains, and the expression of the resistivity is more complex \citep{marchand:16}.
In the two-fluid approximation, the ion inertia is not neglected and the system of equation accounts for both the ions and the neutrals momenta equation. 

The first implementation of ambipolar diffusion in a grid code was presented in \cite{black:82} using a two-fluid approximation. They integrated the induction equation using the ion velocity and treated the collision between neutrals and ions as a friction force in the momenta equations. \cite{maclow:95} present a first 3D implementation of one-fluid ambipolar diffusion in the ZEUS code using a fully explicit scheme as well as CT. They propose to subcycle the magnetic fields evolution in the case when ambipolar diffusion timestep is much shorter than the dynamical one. The first SPH implementation can be found in \cite{hosking:04}, where they used a two-fluid approximation, but their scheme did not preserve the solenoidal constraint. 
\cite{duffin:08} present a one-fluid implementation in the FLASH code using Powell's method \citep{1999JCoPh.154..284P}. \cite{masson:12} also implemented a one-fluid scheme for ambipolar diffusion in RAMSES with the same philosophy as for the Ohmic diffusion. 
A similar implementation using CT, operator splitting, and STS in the ATHENA code can be found in \cite{bai:11} and \cite{chen:14}. 
Last,  \cite{wurster:14} propose a one-fluid implementation for SPH codes, which they implemented with success in PHANTOM.

Other one-fluid implementations in the strong coupling  approximation can be found in \cite{padoan:00} in a 3D explicit, finite difference grid code, \cite{choi:09} in the MHD TVD code  using a flux-interpolated CT scheme \citep{ryu:98,balsara:99}, \cite{christie:17} in the ENZO's Dedner MHD solver. \cite{oishi:06} and \cite{lips:06} presented two simultaneous but independent implementations  in the ZEUS code of a two-fluid solver in the heavy ion approximation.

For star formation applications, most of these methods are usually second-order accurate and at best second-order accurate in time (with a few exceptions though  \citep[e.g.,][]{meyer:14}). Clearly, further work needs to be done to improve the accuracy of resistive MHD solvers. Alongside, multi-fluid approaches are also needed to account for more detailed physics (see Sec. \ref{sec:multifluid}).

\subsection{Hall effect}

The last resistive effect that has been implemented in star formation models is becoming increasingly popular. The Hall effect originates from the drift between ions and electrons, or, more generally the positively and negatively charged particles. The derivation of the generalised Ohm law  (\ref{emf_nimhd}) is not as straightforward as in the case of ambipolar diffusion only. We shall consider the momentum equations of ions and electrons, accounting for the Lorentz force as well as collisions between ions, electrons, and neutrals. If we restrict the resistive effects to the Hall effect only, the induction equation reads
\begin{equation}
\frac{\partial \bf{B}}{\partial t} =    \nabla\times \left[\bf{v} \times \bf{B}  - \frac{\eta_\mathrm{H}}{||{\bf B}^||} {\bf J}\times {\bf B}\right]=  \nabla\times \left[(\bf{v}+{\bf v}_\mathrm{H}) \times \bf{B} \right],
\end{equation}
where we introduce the "Hall speed" ${\bf v}_\mathrm{H} = - \frac{\eta_\mathrm{H}}{||{\bf B}||} {\bf J}$. Contrary to the two previous resistive terms, the Hall effect introduces new waves in the MHD systems: the {\it whistler} waves. The dispersion relation for whistler waves reads \citep{balbus:01}
\begin{equation}
\omega = \pm \frac{\eta_\mathrm{H}k^2}{2}+\sqrt{\left(\frac{\eta_\mathrm{H}k^2}{2}\right)^2+k^2c^2_\mathrm{A}}.
\end{equation}
The whistler waves are dispersive, the phase speed $c_\mathrm{w}=\omega/k$ depends on the wavenumber $k$ and tends to infinity as the wavenumber tends to infinity. Since any discrete formulation cannot extend to infinity, a numerical solver cannot follow the whistler waves with very high frequencies. In practice, the whistler wave speed is chosen to be the one at the grid scale
\begin{equation}
c_\mathrm{w} = \left|\frac{\eta_\mathrm{H}\pi}{2\Delta x}\right|+\sqrt{\left(\frac{\eta_\mathrm{H}\pi}{2\Delta x}\right)^2+c^2_\mathrm{A}}.
\label{cw}
\end{equation}
The time step is then constrained as
\begin{equation}
\Delta t \leq \frac{\Delta x}{c_\mathrm{w}}\approx\frac{\Delta x^2}{\eta_\mathrm{H}}.
\end{equation}

The first way to implement the Hall effect is to use an operator split method, similarly to the Ohm and ambipolar diffusion. A pioneer implementation is presented in \cite{sano:02} in a second-order Godunov finite-difference code \citep{sano:99}, in the local shearing box framework. They use the CT method to update the induction equation with emf computed with the Hall term. The method has been implemented in 3D in ZEUS in \cite{krasnopolsky:11}. \cite{falle:03} presents an implicit scheme for multi-fluid MHD which alleviates the stringent time step conditions, where the time step goes to zero as the strength of the Hall effect increases. 
\cite{osullivan:06} designed a Hall Diffusion Scheme (HDS) in an explicit 3D code for multi-fluid MHD. They split the Hall diffusion in two parts, where the first part uses the STS time integration up to a critical value of the Hall resistivity (i.e. to satisfy the stability condition), and the second part, the HDS,  diffuses the excess Hall resistivity with a different discretisation scheme. The balance between the ambipolar diffusion and the Hall resistivity determines the excess Hall resistivity.  The HDS integrates the induction  equation for the excess Hall using a dimensional splitting which is explicit for the first dimension, i.e. the $x$ component $B_x^{n+1}$ of the magnetic fields is given by $B_y^n$ and $B_z^n$ at time $n$, explicit-implicit for the second dimension, $B_y^{n+1}$ is given by  $B_x^{n+1}$ and  $B_z^{n}$, and finally implicit for third dimension, $B_z^{n+1}$ is given by  $B_x^{n+1}$ and  $B_y^{n+1}$. Both STS and HDS can be subcycled to reach the time step limit given by the hyperbolic system. The HDS scheme has been also implemented in ATHENA by \cite{bai:14} using CT.

An alternative for finite volume methods is to include the Hall effect into the conservative integration scheme. \cite{toth:08} proposed a first implementation in the AMR code BATSRUS \citep{toth:06}. It uses  block-adaptive grids with both explicit and implicit time discretisation, and various methods to control numerical errors in $\nabla\cdot {\bf B}$ (8-waves scheme with diffusion, projection scheme). They achieve spatial second-order convergence by using symmetric slope limiters (like {\it monotonized central}) instead of asymmetric limiters (like {\it minmod}). In addition, they show that the first-order Lax-Friedrich Riemann solver is inconsistent for the Hall MHD equations. This comes from the fact that the whistler wave speed is proportional to the grid spacing, which makes a one order loss in space in the truncation errors of the scheme. The same applies for MUSCL schemes and asymmetric slope limiters which are only first-order accurate. 
\cite{lesur:14} built upon \cite{toth:08} a Hall MHD solver in PLUTO for protoplanetary disc evolution studies. They implement a HLL solver to estimate the flux and then update the magnetic fields using CT. When written in a conservative form, the Hall induction equation reads
\begin{equation}
\frac{\partial {\bf B}}{\partial t} = \nabla \cdot\left[{\bf vB} - {\bf Bv} - \frac{\eta_\mathrm{H}}{||{\bf B}||}({\bf JB} - {\bf BJ})\right].
\end{equation}
The current ${\bf J}=\nabla\times{\bf B}$ involves spatial derivatives of the magnetic fields, so that the flux depends on the conserved quantities, and on its derivative. This leads to a ill-defined Riemann problem. \cite{lesur:14} solved this issue by assuming that {\bf J} is an external parameter, which is computed at the cell interfaces using the predicted states. The left and right fluxes take this unique value as an  input, with the predicted {\bf B}. The Godunov fluxes are then computed using a whistler-modified HLL solver, which accounts for the whistler waves speed (\ref{cw}) in the characteristics speeds.  

Last, \cite{marchand:18} recently extended the \cite{lesur:14} implementation to the CT scheme of RAMSES, which uses 2D Riemann solver to integrate the induction equation. They designed a 2D whistler-modified HLL solver to estimate the emf on cell border, assuming a uniform current in the 2D Riemann solver, again computed from the predicted states.

In SPH codes, the Hall effect has been implemented and used with success for star formation applications in \cite{wurster:16}, \cite{tsukamoto:17} and \cite{price:17}. The standard method is based on an operator splitting and either a full explicit or a STS scheme based on the implementation for ambipolar diffusion \citep{wurster:14}. 

To date, only \cite{krasnopolsky:11}, \cite{wurster:16}, \cite{tsukamoto:17} and \cite{marchand:18} have applied their methods for protostellar collapse. While the SPH methods seem to lead to a relatively accurate conservation of the angular momentum, the grid-based methods using CT are suffering from severe non-conservation of the angular momentum. The origin of this non-conservation is unclear. \cite{krasnopolsky:11} invoked Alfv\'en waves going out of the simulation volume, while \cite{marchand:18} did not find evidence of this. Instead, they propose that the non-conservation takes place after the accretion shock formation. Shocks indeed generate strong gradients and thus large Hall velocities. Further investigation is clearly needed to cope with this fundamental issue.

\subsection{Full resistive implementation}

Currently, a handful of 3D codes benefit of a full implementation of the three resistive terms: PLUTO \citep{lesur:14}, ATHENA \citep{bai:14}, RAMSES \citep{masson:12,marchand:18}, the SPH code by \cite{tsukamoto:17}, PHANTOM \citep{price:17}, ZeusTW \citep{li:11}, and GIZMO \citep{hopkins:17}. A very recent full implementation for solar physics can be found in \cite{gonzalez-Morales:18} in the MANCHA3D grid code. It combines STS for ambipolar and ohmic diffusion, and HDS for the Hall effect. 

Currently, the ATHENA, PLUTO, RAMSES, and ZeusTW implementations  rely on CT algorithms for the resistive MHD equations integration. All other works use divergence cleaning algorithms.  Given the variety of the methods developed in the literature, a quantitative comparison (in standard tests as well as in star formation applications) of all these implementations would be welcome in the coming years.

Last but not least, it is not clear which non-ideal process dominates over the others. The physical conditions in star-forming regions are so wide that there is room for regions where all three effects dominate. In addition, every effect needs to be tested and quantified before any conclusions on the relative importance of each of these effects  \citep[see the review by][]{wurster:18b}.

\subsection{Multifluid approach}
\label{sec:multifluid}

Multifluid approaches are designed to describe multiphase or multifluid flows. In the context of star formation, different fluids or phases are at stack: neutrals, ions, electrons, and dust grains. A few attempts have been made to describe all or a part of these components using multi-fluid approach. In this section, we consider only the approaches which do account for $N$ momentum equations, where $N$ is the number of phases or fluids. For instance, a popular two-fluid framework considers ions and neutrals and describes the friction between these two fluids, i.e. the ambipolar diffusion. 
 \cite{toth:94} presented a first 2D implementation of two-fluid ions and neutrals dynamics, which is based on a flux-corrected transport finite difference method. \cite{inoue:08} propose an unconditionally stable numerical method to solve the coupling between two fluids ions-neutrals (frictional forces/heatings, ionization, and recombination, \citep{draine:86} for an ISM gas subject to the thermal instability. They split time integration as follows: 1/ the ideal hydrodynamical part for neutrals, 2/ the ideal MHD part for ions. The first part is integrated using a second-order Godunov scheme. The ideal MHD part is solved in two steps, similarly to what is done for one-fluid ideal MHD solvers \citep[e.g.][]{1992ApJS...80..791S}: the magnetic pressure terms are solved using a second-order Godunov method, and the magnetic tension term and the induction equation are solved using the method of characteristics for Alfv\`en waves with the CT algorithm. 
\cite{tilley:08}  present a semi-implicit method for two-fluid ion-neutral ambipolar drift in the RIEMANN code \citep{2001JCoPh.174..614B,balsara:04}. Their scheme is second-order accurate in space and time and uses a mixed implicit-explicit formulation to deal with strong friction regimes. 
\cite{pinto:08} consider a system composed of three fluids: positively-charged particles, negatively-charged
particles, and neutrals. The charged particles include ions, electrons, as well as charged dust grains, depending on the main charge carrier. The self-consistent set of equations they designed remains to be implemented in 3D codes.

Last but not least, a complete multi-fluid approach should include dust grains dynamics on top of the neutrals, ions, and electrons.  \cite{falle:03} presented a scheme for multi-fluid hydrodynamics in the limit of small mass densities of the charged particles, i.e., where the inertia of the charged particles are neglected. This is a similar  to the heavy ion approximation but for neutrals, ions, electrons, and dust grains of various sizes. The HD equations are first integrated for the neutrals using a second-order Godunov scheme, and then the induction equation and the charged species velocities are updated using an implicit scheme. This scheme captures the three non-ideal effects and can deal with regime where the Hall effect dominates. We note that similar approaches were introduced in \cite{ciolek:02} and \cite{osullivan:06}.

The most complete work on multi-fluid designed for star formation was presented in \cite{kunz:09}. They considered a six-fluid set of equations (neutrals, electrons, molecular and atomic ions, positively charged, negatively charged, and neutral grains) and implemented it in a 2D version of the radiation-MHD ZEUS-MP code \citep{hayes:06}. They modified the MHD solver to account for non-ideal effects (Ohmic dissipation and ambipolar diffusion). The abundances of the 6 species are calculated using a reduced chemical-equilibrium network. They applied their methods to protostellar collapse in \cite{kunz:10}. Unfortunately, the CPU cost of such a complete set of physics is prohibitive for 3D models, and no work has taken over since this first application.

\subsection{What is next? Dust and gas mixture dynamics}

Currently, numerous studies report on the effect of non-ideal MHD in the star formation process. The wide majority of these works use the single-fluid approximation, assuming furthermore that the dust grains are perfectly coupled to the gas via collisions. Nevertheless, dust grains are observed to cover a wide range of sizes, from nanometer to micrometer in the ISM \citep{mathis:77} and even millimeter in protostellar disks environments \citep{chiang:12}. Generally, it assumes that dust grains of different sizes coexist and follow a power-law dust distribution. 
Grains react to changes in the gas velocity via a drag force. For a given dust grain of mass $m_\mathrm{g}$, the force is
\begin{equation}
{\bf F}_\mathrm{drag}=-\frac{m_\mathrm{g}}{t_\mathrm{s}}({\bf v}_\mathrm{g}-{{\bf v}}),
\end{equation}
where ${\bf v}_\mathrm{g}$ is the dust grain velocity, ${\bf v}$ the gas velocity, and $t_\mathrm{s}$ the stopping time, i.e., the characteristic decay timescale for the dust grain velocity relative to the gas. 
In star forming regions, the mean free path of the gas molecules is larger than the dust particles radius and the stopping time depends linearly on the dust grain size $t_\mathrm{s} \sim s$ \citep{epstein:24}. The Stokes number ${\rm St}\equiv t_\mathrm{s}/t_\mathrm{dyn}$ characterizes the coupling of the dust grain relative to the gas, where well coupled dust grains follow ${\rm St}<1$. $t_\mathrm{dyn}$ represents a characteristic dynamical time, e.g., the eddy turbulent crossing time for molecular clouds or the orbital time for prostostellar discs. The Stokes number thus depends on the dust grain size so that part of the dust grains of a given dust size distribution will be coupled to the gas and the other part will be decoupled.

\begin{figure}[h!]
	\begin{center}
		\includegraphics[width=17cm]{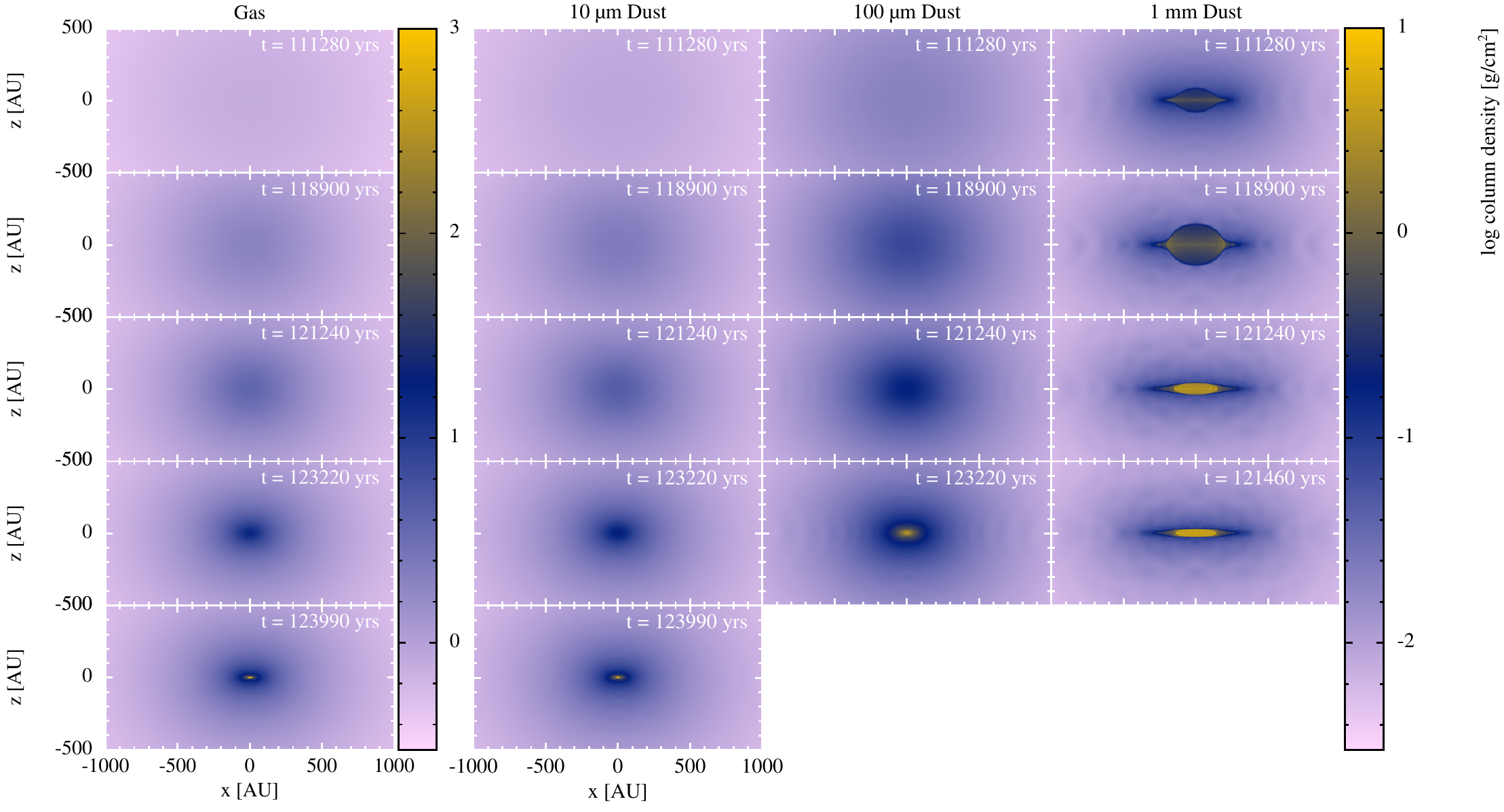}
	\end{center}
	\caption{SPH calculations of dust dynamics during the collapse of a 1~\msun~dense core. The left column shows the gas column density (edge-on view) and from left to right, the three other columns show the dust column density for dust grain sizes of 10~$\mu$m, 100~$\mu$m, and 1~mm. Large dust grains decouple from the gas and settle in the mid-plane faster into a large dusty disc with larger dust-to-gas ratio than in the envelop. Figure adapted from \cite{bate:17} with permission from the authors. \label{fig:bate}}
\end{figure}

Several implementations of dust and gas mixtures have been developed in the literature to deal with the various Stokes regime. For ${\rm St}>1$, the bi-fluid formalism seems to be the most adapted \citep[e.g.][]{bai:10,meheut:12,laibe:12,loren:14,booth:15}. However, bi-fluid algorithms encounter sever limitations in the strong drag regime (small stopping time), where prohibitive spatial and time discretisations may be required \citep{laibe:12}. In recent years, \cite{laibe:14} proposed a monofluid formalism for the dust and gas mixture which is well adapted to regimes with ${\rm St}\leq 1$. This monofluid formalism has been implemented with success in SPH and grid-based codes in the past five years \citep{laibe:14b,lebreuilly:19}. Last, \cite{price:15} and \cite{lin:17} present  methods for simulating the dynamics of small grains with only one additional equation on the dust concentration on top of the Euler equations. These last methods have been recently implemented with success in PHANTOM \citep{price:17}, as well as in PLUTO \citep{chen:18} and in RAMSES \citep{lebreuilly:19}.

Recent works have emphasized the possible decoupling of micrometer dust grains in molecular clouds \citep{hopkins:16,tricco:17} as well as of millimeter grains in collapsing dense cores \citep[][see Fig.~\ref{fig:bate}]{bate:17}. This decoupling is at the origin of a dynamical sorting which changes the dust size distribution. Other mechanisms may also affect the size distribution: coagulation, fragmentation, sublimation, etc.  Further work should investigate the relative importance of each of these processes on the dust size distribution. In addition, the shape of the dust size distribution plays a critical role in the calculations of the resistivity as well as of the opacity. We can anticipate that coupled dust and gas dynamics will consider the evolution of the dust size distribution, coupled  to non-ideal MHD and RHD algorithms.  Last but not least, all current works on this field only account for the hydrodynamical (the pressure gradient) and gravitational forces. Adding the dynamics of charged dust grains is the natural next step.

\section{Radiative transfer: numerical techniques}
\label{Sec:rad_trans}

\subsection{A poor's man approach: the polytropic equation of state}

Radiative transfer plays a fundamental role in star formation and cannot be ignored, even for a review on
magnetic fields. The importance of radiation fields is two-fold. First, when the molecular cores
collapse and reach a high enough density, the dust opacity becomes large enough to absorb the cooling
radiation and the gas becomes optically thick. The traditional value for the dust opacity is
\begin{equation}
\kappa_{\rm dust} \simeq 0.1~{\rm cm}^2/{\rm g}~~~{\rm at}~~~T \simeq 10{\rm K}.
\end{equation}
Requiring that the Jeans length of the gas becomes optically thick to infrared radiation leads to 
\begin{equation}
\rho \kappa_{\rm dust} \lambda_{\rm J} > 1~~~{\rm or}~~~\rho > \rho_{\rm crit} = \frac{G}{\pi \kappa_{\rm dust}^2 c_{\rm s}^2} \simeq 10^{-14}~{\rm g/cm}^3.
\end{equation}
Many MHD simulations use this critical density to define a polytropic equation of state of the form
\begin{equation}
T(\rho) = 10{\rm K} \left( 1  + \left( \frac{\rho}{\rho_{\rm crit}}\right)^{2/3}\right) ,
\end{equation}
to capture the transition from an optically thin, isothermal gas at low density to an optically thick,
adiabatic gas at high density. Although this simplified approach is very useful for many already expensive simulations of 
collapse of turbulent clouds, this does not model properly the physics of radiative transfer \citep[e.g.][]{commercon:10}. 
More importantly, once protostars have formed, higher energy radiation will emerge from the accretion shocks on the surface
of the protostars, or from the main sequence stars after they ignited nuclear reactions. This high energy radiation in the optical 
or UV bands will then interact with the parent cloud, providing {\it radiation feedback} and influence the formation of the next generation
of stars \citep{2007ApJ...656..959K,Offner:2009jb,2013ApJ...766...97M}. 
In conclusion, radiation plays a central role during the formation and the evolution of the first Larson core (see Sec.~\ref{Sec:2nd_col}),
but later on acts as a self-regulating mechanism for star formation.

\subsection{Ray tracing and long characteristic methods}

The radiative transfer equation written in full generality using the radiation specific intensity $I_\nu({\bf x},{\bf n},t)$
\begin{equation}
\frac{1}{c}\frac{\partial}{\partial t} I_{\nu} + {\bf n} \cdot \nabla I_{\nu} = j_{\nu} - \alpha_{\nu} I_{\nu}.
\end{equation}
{bf where $j_\nu$ is the emission coefficient and $\alpha_\nu$ is the absorption coefficient.
Note that the radiative transfer equation is written here in the laboratory frame. 
The absorption and emission coefficients are however defined in the comoving frame. 
Properly introducing the relativistic corrections to the radiative transfer equation and its various moments
is a very important aspect of the problem we will only briefly touch upon in this review.

The most natural numerical technique to solve the radiative transfer equation is {\it ray-tracing}.
The idea is to shoot a discrete set of light rays from a point source, solving the previous equation along the path of the
light ray using the curvilinear coordinate $s$ defined as ${\rm d}s = c {\rm d}t$. The equation of radiative transfer
can be written as a Lagrangian time derivative following the trajectory of the photons as
\begin{equation}
\frac{{\rm d}}{{\rm d}s} I_{\nu} = j_{\nu} - \alpha_{\nu} I_{\nu}.
\end{equation}
The light ray is discretised along its path, matching the underlying gas distribution, whether it is a grid or
a set of SPH particles. The difficulty is then to choose the appropriate number of rays, so that the angular distribution
of the radiation field is properly sampled, but also so that enough light rays intersect the gas elements. 
In \cite{2002MNRAS.330L..53A}, an adaptive method was proposed, that splits or merges light rays adaptively,
using the Healpix tessellation to sample the sphere. This method was first developed for the ENZO code \citep{2011MNRAS.414.3458W}
and used with success to describe photo-ionization regions at the Epoch of Reionization \citep{wise:12}, as well as in turbulent molecular clouds \citep{shima:17}. 
A similar scheme based on the Monte-Carlo approach has been developed for SPH in \cite{2008MNRAS.386.1931A} and \cite{2008MNRAS.389..651P}.

A similar approach is provided by the long characteristics method. This method can describe light rays emanating from 
a single point source like the ray-tracing method, but also diffuse radiation by adopting for each direction 
a set of parallel rays whose spacing matches the local grid resolution. This method has been implemented in the FLASH AMR code
by \cite{2006A&A...452..907R}, with however an important addition we will discuss later. Diffuse radiation using long characteristics
was included later by \cite{buntemeyer:16} in the FLASH code, with an emphasis of describing both the optically thin and optically thick
regimes. These methods exploit the particular property of the AMR implementations of FLASH and ENZO, namely large rectangular patches
of various sizes and resolution. The long characteristic method has been developed for the graded octree AMR code RAMSES only recently
by \cite{2018arXiv180905541F}. 

\subsection{Monte Carlo and short characteristic methods}

Although long characteristics offer the advantage of solving the exact attenuation along light rays, an especially accurate approach
for point sources, they are tricky to parallelise because light rays propagate over multiple MPI domains. An alternative technique is the
short characteristics method, for which light rays as short as the hydrodynamics cell size are considered. The radiation
intensity from neighboring light rays is interpolated between neighboring cells, avoiding the use of long characteristics. 
\cite{2006A&A...452..907R} use an hybrid long- and short-characteristics method in the FLASH code to overcome the difficulty to 
parallelise the long characteristics method. The short characteristic method also offers better scaling properties.
The short characteristics method has also been used in the ATHENA code to solve the stationary radiative transfer equation by 
\cite{2012ApJS..199....9D} and in the C2-ray code for ionization problems \citep{2006NewA...11..374M}.

A third method which shares with the two previous ones the property of being accurate and able to capture complex radiation geometry is the
Monte Carlo method. The Monte Carlo approach samples the radiation field with photon packets carrying information such as angle of propagation
and frequency. This is as close as one can be from true photons propagating in the fluid and allows for complex scattering processes
to be included in the models \citep{2003MNRAS.340.1136E,2008MNRAS.386.1931A,2011MNRAS.416.1500H,2015ApJS..217....9R}. 
The method can be however quite expensive, as large particle numbers are required to avoid
that the Poisson noise dominates the computed signal. 
In the diffusion limit, the mean free path becomes much smaller than the length scale of the system and the number of
interactions becomes prohibitively large: It scales as the square of the optical depth \citep{2009MNRAS.397.1314N}. 
It has also been mostly developed for stationary problems but many codes able to deal with non-stationary problems 
are now emerging \citep{2009MNRAS.397.1314N,2011MNRAS.416.1500H,2016MNRAS.461.3542L}.

\subsection{Two moments methods}

The common aspect in these three methods is that they can be quite expensive. This is probably not true for the short characteristics method,
although the number of angular domains adopted plays a role in the final accuracy of the solution, and affects the cost. An alternative
approach, called moment-based radiative transfer, can potentially solve this cost problem by integrating the radiative transfer 
equation over the angles. We obtain the radiation energy and the radiation momentum equation as
\begin{equation}
\frac{\partial}{\partial t}E_{\nu} + \nabla \cdot {\bf F}_{\nu} = 4 \pi j_\nu - \alpha_\nu c E_{\nu},
\end{equation}
\begin{equation}
\frac{1}{c}\frac{\partial}{\partial t}{\bf F}_{\nu} + c \nabla \cdot {\mathbb P}_{\nu} = - \alpha_\nu {\bf F}_{\nu},
\end{equation}
where $E_{\nu}$ is the radiation energy, ${\bf F}_\nu$ is the radiation flux and ${\mathbb P}={\mathbb D} E_{\nu}$
is the radiation pressure tensor and ${\mathbb D}$ is the Eddington tensor.
For slowly varying spectral absorption and emission coefficients, for which one has
\begin{equation}
\nu \frac{\partial \alpha_\nu}{\partial \nu} \ll \alpha_\nu~~~{\rm and}~~~\nu \frac{\partial j_\nu}{\partial \nu} \ll j_\nu,
\end{equation}
one can write relativistic correction to first order 
in $(v/c)$ as
\begin{equation}
\alpha_\nu({\bf n}) = \alpha^0_\nu - ({\bf n}\cdot {\bf v}/c) \alpha^0_\nu,
\end{equation}
\begin{equation}
j_\nu({\bf n}) = j^0_\nu + 2 ({\bf n}\cdot {\bf v}/c) j^0_\nu,
\end{equation}
where superscript $0$ refers to quantities measured in the comoving frame \citep[see][for a detailed discussion]{Mihalas:1982dp,1984oup..book.....M,2007ApJ...667..626K}. 
Injecting this in the radiative 
transfer equation and taking the moments over the angles leads to new terms accounting for the relativistic corrections
\begin{equation}
\frac{\partial}{\partial t}E_{\nu} + \nabla \cdot {\bf F}_{\nu} = 4 \pi j^0_\nu - \alpha^0_\nu c E_{\nu} + \alpha^0_\nu {\bf F}_\nu \cdot \frac{{\bf v}}{c},
\end{equation}
\begin{equation}
\frac{1}{c}\frac{\partial}{\partial t}{\bf F}_{\nu} + c \nabla \cdot {\mathbb P}_{\nu} = - \alpha^0_\nu {\bf F}_{\nu} + \alpha^0_\nu {\mathbb P}~{\bf v} + 4\pi j^0_\nu \frac{{\bf v}}{c}.
\end{equation}
Since $E_\nu$ and ${\bf F}_\nu$ are still expressed in the laboratory frame but $\alpha^0_\nu$ and $j^0_\nu$ are expressed in the comoving frame, 
these equations are often called {\it mixed frame} radiation moments equations. We can also Lorentz-transform the radiation moments from the laboratory
frame to the comoving frame, which gives, to first order in $(v/c)$
\begin{equation}
E^0_{\nu} = E_{\nu} - \frac{2}{c}{\bf F}_{\nu} \cdot \frac{{\bf v}}{c},
\end{equation}
\begin{equation}
{\bf F}^0_{\nu} = {\bf F}_{\nu} - E_\nu {\bf v} - {\mathbb P}~{\bf v} .
\end{equation}
Injecting these relations {\it only in the right-hand side} of the moments equations and neglecting all $(v/c)$ terms gives the simpler form
\begin{equation}
\frac{\partial}{\partial t}E_{\nu} + \nabla \cdot {\bf F}_{\nu} = \left( 4 \pi j^0_\nu - \alpha^0_\nu c E^0_{\nu} \right) - \alpha^0_\nu {\bf F}^0_\nu \cdot \frac{{\bf v}}{c},
\end{equation}
\begin{equation}
\frac{1}{c}\frac{\partial}{\partial t}{\bf F}_{\nu} + c \nabla \cdot {\mathbb P}_{\nu} = - \alpha^0_\nu {\bf F}^0_{\nu} + \left( 4\pi j^0_\nu - \alpha^0_\nu c E^0_{\nu} \right) \frac{{\bf v}}{c},
\end{equation}
where all right-hand side terms are now expressed in the comoving frame. 
In practice, after the radiation transport step, encoded in the left-hand side, one then converts the laboratory frame radiation variables
into the comoving frame and solve the thermo-chemistry encoded in the right-hand side in the comoving frame. Local thermodynamical equilibrium 
conditions can then be reached in the comoving frame, eventually reaching the diffusion limit (see below). At the end of the thermo-chemistry
step, one finally converts back the radiation variables into the laboratory frame before entering the next transport step \citep[see][for implementation details]{2015MNRAS.449.4380R}.

These equations are quite handy because of their simplicity: We have no angular dependence anymore. There is a catch however, since everything
depends on the Eddington tensor, for which we have no equation at this order of the moment's hierarchy.
At this point, two different strategies have been explored in the star formation literature: 1- compute an accurate Eddington tensor
by solving exactly the {\it stationary} radiative transfer equation using one of the methods presented above (short or long characteristics, 
ray-tracing or Monte Carlo), 2- compute an approximate form of the Eddington tensor based on a particular closure of the moment's hierarchy.

The first approach is often referred to a the Variable Eddington Tensor (VET) method. 
In \cite{2001NewA....6..437G}, the authors developed a moment-based solver coupled to a cosmological hydrodynamics code, called
Optically Thin Variable Eddington Tensor or OTVET. A similar technique has also been implemented for the SPH code GADGET \citep{2011MNRAS.415.3731P}.
The idea is to solve the stationary radiative transfer equation in the full domain 
assuming an optically thin medium, so that the problem boils down to a collection of $1/r^2$ sources combined together. This can be achieved
efficiently using any gravity solver. Once the corresponding Eddington tensor is computed, the moments equations are solved using a finite difference
scheme. In \cite{2012ApJS..199...14J}, on the other hand, 
the authors developed a moment-based solver for the ATHENA MHD code, using VET together with the short characteristic method of \cite{2012ApJS..199....9D} 
to compute the Eddington tensor. Obviously, this leads to significantly more accurate results in optically thicker environments. 
An intermediate solution has been implemented in the TreeCol code \citep{2012MNRAS.420..745C}, for which a gravity tree code is used to collect column 
densities between particles.

The second approach relies on simple but approximate models for the Eddington tensor. The simplest one, called M0, assumes
the Eddington tensor is isotropic, with
\begin{equation}
{\rm M0~closure:}~~~{\mathbb D} = \frac{1}{3} {\mathbb I},
\end{equation}
as it is the case for example in the diffusion limit (see below). A slightly more elaborate model is called the M1 closure \citep{dubroca:99,ripoll:01}.
It assumes that the radiation intensity can be fitted by a Lorentz-boosted Planckian distribution, in other words an ellipse in 
angular space. The parameters of this distribution are determined by matching the radiation energy and the radiation flux. This leads to 
\begin{equation}
{\rm M1~closure:}~~~{\mathbb D} = \frac{1-\chi}{2} {\mathbb I} + \frac{3 \chi-1}{2} {\bf n} \otimes {\bf n}.
\end{equation}
This closure captures the optically thick regime, but also optically thin conditions,
in case there is one dominant source of radiation. Indeed, in this case, one gets the free-streaming regime for radiation with ${\mathbb D} \simeq {\bf n} \otimes {\bf n}$.
The unit vector ${\bf n}$ and the parameter $\chi$ are determined using the radiation energy and 
the radiation flux as
\begin{equation}
{\bf n}=\frac{{\bf F_\nu}}{\left| {\bf F_\nu} \right| }, ~~~\chi(f) = \frac{3+4f^2}{5+2\sqrt{4-3f^2}}, ~~~f=\frac{\left| {\bf F_\nu} \right| }{cE_\nu}
\end{equation}
A particularly interesting property of the M1 model is that it leads to a system of hyperbolic conservation laws, that can be numerically integrated
using Godunov schemes. 
The M1 model has also some major caveats, such as radiation shocks, in case multiple strong sources are present, or incorrect radiation geometry, 
in case of sharp shadows.
This model was introduced for the first time in astrophysics by \cite{2007A&A...464..429G} in the HERACLES code. It was then adapted 
to cosmic reionization by \cite{2008MNRAS.387..295A} in the ATON code, and ported to GPU into the RAMSES-CUDATON code \citep{2010ApJ...724..244A}, used now routinely for 
simulations of the Epoch of Reionization (EoR) \citep[e.g.][]{2016MNRAS.463.1462O}. Later on, the M1 method has been ported to the AMR code RAMSES, leading 
to the design of the RAMSES-RT solver \citep{2013MNRAS.436.2188R,2015MNRAS.449.4380R}. 
The M1 method has also been ported to ATHENA by \cite{2013ApJS..206...21S} and recently to AREPO by \cite{2018arXiv180401987K}, 
the latter based on a new higher-order implementation of the M1 closure. 

\subsection{Flux-limited diffusion methods}

A definitive advantage of moment-based radiative transfer methods is the possibility to model both optically thin and optically thick regimes.
For optically thick conditions, the radiation field follows the diffusion limit of the radiative transfer equation. In that limit,
the radiation field {\it in the comoving frame} is quasi-isotropic, so that ${\mathbb D} \simeq \frac{1}{3}{\mathbb I}$ and the radiation flux can be 
approximated by
\begin{equation}
{\bf F}^0_{\nu} \simeq - \frac{c}{3 \alpha^0_\nu} \nabla E^0_\nu.
\end{equation}
Since now the flux in the comoving frame is a direct function of the energy density in the comoving frame, we do not need the 
radiation flux equation anymore and we are left with only the radiation energy equation. The Lorentz transform can be written
to leading order as
\begin{equation}
E_{\nu} \simeq E^0_{\nu} 
\end{equation}
\begin{equation}
{\bf F}_{\nu} \simeq -\frac{c}{3 \alpha^0_\nu} \nabla E^0_\nu + \frac{4}{3}E^0_\nu {\bf v}.
\end{equation}
Injecting these into the radiation energy equation leads to a fully covariant formulation in the diffusion limit
\begin{equation}
\frac{\partial}{\partial t}E^0_{\nu} + \nabla \cdot \left( \frac{4}{3} E^0_\nu{\bf v} \right) - \nabla \cdot \left( \frac{c}{3 \alpha^0_\nu} \nabla E^0_\nu \right) 
= \left( 4 \pi j^0_\nu - \alpha^0_\nu c E^0_{\nu} \right) + \frac{1}{3} \nabla E^0_\nu \cdot {\bf v},
\end{equation}
which can be written in the familiar form
\begin{equation}
\frac{\partial}{\partial t}E^0_{\nu} + \nabla \cdot \left( E^0_\nu{\bf v} \right) + \frac{1}{3} E^0_\nu \nabla \cdot {\bf v} =
 \nabla \cdot \left( \frac{c}{3 \alpha^0_\nu} \nabla E^0_\nu \right) + \left( 4 \pi j^0_\nu - \alpha^0_\nu c E^0_{\nu} \right).
\end{equation}
This is the radiation energy equation in the diffusion limit using only comoving radiation variables. If one wants to extend
this equation outside its validity range (high optical depth), then one can use Flux Limited Diffusion (FLD),
for which the flux function is modified as 
\begin{equation}
{\bf F}^0_{\nu} \simeq - \frac{c \lambda(R)}{\alpha^0_\nu} \nabla E^0_\nu~~~{\rm and}~~~R = \frac{\left| \nabla E^0_\nu\right|}{\alpha^0_\nu E^0_\nu},
\end{equation}
where $\lambda(R)$ is the flux limiter, a function that has to connect the diffusion limit with $\lambda \simeq 1/3$ for $R \simeq 0$
to the free-streaming regime where $\lambda \simeq 1/R$ for $R \rightarrow +\infty$. A possible form for the flux limiter has been proposed by
\cite{minerbo:78} and reads
\begin{eqnarray}
\lambda = \frac{2}{3+\sqrt{9+12R^2}}~~~&{\rm for}&~~~0\le R \le \frac{3}{2},\\
\lambda = \frac{1}{1+R+\sqrt{1+2R^2}}~~~&{\rm for}&~~~\frac{3}{2}\le R \le +\infty.
\end{eqnarray}
while the \cite{1984JQSRT..31..149L} flux limiter reads
\begin{eqnarray}
\lambda = \frac{1}{R} \left( \coth R - \frac{1}{R}\right)
\end{eqnarray}

Historically, FLD has been the first method implemented in radiation hydrodynamics codes. Here again, the ZEUS 2D code
was a precursor \citep{1992ApJS...80..819S}. For AMR, the ORION code was developed specifically for FLD in collapsing 
star forming core by \cite{2007ApJ...667..626K}, using the mixed frame formulation outlined above or in \cite{Mihalas:1982dp}. 
A version of the RAMSES
code with FLD was developed later by \cite{2011A&A...529A..35C}, with a particular emphasis on adaptive time stepping
together with implicit time integration described in \cite{2014A&A...563A..11C}. Because of its simplicity, FLD has been used
for many detailed studies of gravitational collapse with the effect of radiation included. The current frontier for FLD
is probably the multigroup treatment of radiation, allowing for more accurate  models of the full spectral energy distribution
\citep{2008JCoPh.227.2154S,2013ApJS..204....7Z,2015A&A...578A..12G}. In the context of multifrequency radiative transfer, 
hybrid solutions have also been proposed, with ray-tracing techniques dedicated to the high-energy radiation field, and
FLD for the infrared, dust-absorbed radiation field \citep{2010A&A...511A..81K,klassen:14}.

FLD has also been developed for SPH by  \cite{2004MNRAS.353.1078W}, \cite{2005MNRAS.364.1367W} and \cite{2007ApJ...661L..77m}, where most of the difference
with the many grid-based versions lies in the radiation energy gradient that uses the SPH kernel. A simplified version based on a local
estimate of the column density has also been proposed by \cite{stamatellos:07} which avoids entirely the need for an implicit radiation solver.
This method can be considered as an intermediate solution between the polytropic equation of state
that we introduced at the beginning of this section and FLD. 
The main caveat here is the complete lack of any proper transport mechanism for radiation.
In a similar spirit, \cite{dale:07} also developed an approximate method for ionizing radiation, 
with Stromgren sphere iteratively grown around each star to locate photo-ionized, photo-heated gas.


\section{Second collapse}
\label{Sec:2nd_col}

From first principles, the ultimate goal of star formation is the formation of the protostar itself. The major difficulty sits in the huge  dynamical range in  physical and temporal scales that have to be described: protostars with radii of about 1~\rsun~ form within dense cores of sizes $\simeq 0.1$~pc. So there are two ways to deal with protostars in star formation. The first one is to try to follow the collapse down to the stellar scales with the best numerical model accounting for the complex physical processes at stack combined with a very high numerical resolution at a cost of very short horizon of predictability. The second is the opposite: the physics and numerical resolutions are degraded, but the models are integrated over longer dynamical timescales to study the impact of protostellar feedback on the ISM dynamics. In the next three sections, we review the work that has been done using these two approaches, as well as attempts to account for protostellar evolution in large-scale models.

\subsection{Historical work and 1D studies}

Contemporary numerical star formation studies started in the late 60s with the outstanding pioneering work of \cite{larson:69} who first computed numerically the collapse of a dense core down to the formation of the protostar. He used a 1D spherically symmetric model, accounting for coupled gas dynamics and radiative transfer in a modified Eulerian scheme. Larson identified two distinct stages during the protostellar collapse, called the first and the second collapse. Each of these two stages are followed by the formation of a hydrostatic object, commonly referred to as the first and second Larson cores in post \cite{larson:69} studies. 
Larson's work established an empirical evolutionary sequence as follows. Dense cores first collapse isothermally because dust thermal emission is very efficient at radiating away the gas compression energy \citep[dust and gas are thermally coupled within dense cores, see for instance][]{galli:02}. The first collapse stops at  densities of the order of $\simeq 10^{-13}$~g~cm$^{-3}$ at which dust grains become opaque to their own radiation so that the gas begins to heat up quasi-adiabatically. This is the formation of the first Larson core. The first collapse lasts roughly a free-fall time. The first core accretes matter and its temperature increases up about 2000K where H$_2$ dissociation starts. This endothermic reaction enables the gas to behave as a fluid with an effective polytropic index smaller than the critical value $4/3$ for collapse. The second collapse thus starts at typical densities of $\simeq 10^{-8}$~g~cm$^{-3}$ until stellar densities are reached within $\simeq100$~yr to form the second core, i.e., the protostar.  The typical properties of first Larson cores at the start of the second collapse are a size $\simeq10$~AU, a mass $\simeq0.01$~\msun, and a lifetime $\simeq1000$~yr. We note that first Larson cores are predicted by theoretical and numerical studies, but there is no observational confirmation of such objects even though a few sources are considered as good candidates \citep{tsitali:13,gerin:15,maureira:17}.

Since Larson's work numerous 1D calculations using spherical symmetry have been conducted. We note the work in the 80s by \cite{stahler:80} and \cite{winkler:80} in which Larson predictions were confirmed to establish the current empirical evolution sequence of collapsing low-mass dense cores. This evolutionary picture is still currently the commonly admitted scenario for  low-mass star formation. For massive star formation, recent work indicates that first cores may not have time to form because of the large accretion rates \citep{bhandare:18}. Note that in the reminder of the section dedicated to second collapse, we mention only the work which is consistent with the Larson evolutionary picture, i.e., where the gas thermal and chemical budgets are taken into account, and we do not mention the work that has been done  using an isothermal approximation. 

Modern studies began with the calculations performed in \cite{masunaga:00} who incorporated a more accurate radiation transport scheme as well as a realistic gas equation of state which accounts for H$_2$ dissociation. \cite{masunaga:00} integrated their models throughout the Class 0 and Class I phases up to an age of $1.3\times 10^5$~yr. Nowadays, 1D models are still used either as a first step towards describing more accurately the physics of the protostellar collapse, e.g., with multigroup radiative transfer \citep{vaytet:13,vaytet:14} or with a view to provide quantitative predictions for the first and second Larson core properties \citep{vaytet:17,bhandare:18}. These models are not pushed in time as far as the ones by \cite{masunaga:00}.

\subsection{From 1D to 3D}

Pioneered studies acknowledged up front that 1D spherical models were limited because the effect of rotation, turbulence, and magnetic fields cannot be taken into account. Three decades after Larson work, the first 3D numerical simulation able to describe the formation of the protostar was performed by \cite{bate:98} using SPH. The model accounted for initial rotation, but magnetic fields were neglected and radiative transfer was crudely mimicked by a piecewise polytropic equation of state. It is worth to notice at this point that grid-based codes took some time to reproduce \cite{bate:98} results for almost one decade. SPH is indeed naturally well-suited to collapse problems thanks to its Lagrangian nature while innovative techniques, such as AMR and nested grids, were not mature enough to capture the entire dynamical range and to cover eight orders of magnitude in physical length given the computer capabilities at that time. As mentioned in the Sec.~\ref{imhd}, ideal MHD was introduced early in protostellar collapse models using grid-based codes \citep[e.g.,][]{dorfi:82,phillips:85}.  

The first calculations of second collapse incorporating magnetic fields were thus done on a grid-based code by \cite{tomisaka:02} using a nested grid with cylindrical coordinates assuming axisymmetry \citep{tomisaka:93}. \cite{tomisaka:02} code solved the ideal MHD equations using a second-order ``monotonic scheme'' \citep{vanLeer:77,norman:86}, a constrained transport scheme for the induction equation \citep{1988ApJ...332..659E}, as well as an angular momentum preserving scheme. To capture the huge range in spatial scales, \cite{tomisaka:02} used up to 16 levels of refinement.  He found that two different types of outflows were launched at the first and second cores scales. He suggested that the second core outflow corresponds to the optical jets observed in young stellar objects (YSOs), while the first core outflow corresponds to the molecular bipolar outflow. The first full 3D MHD models were presented in \cite{machida:06} using nested grid with Cartesian coordinates and 21 levels of refinement. They compared the results obtained with resistive (Ohmic diffusion) and ideal MHD. They confirmed the results of \cite{tomisaka:02} even with non-ideal MHD included. After \cite{machida:06} work, the numerical challenge of forming a protostar in 3D with MHD was faced, but the horizon of predictability remained very short since they had to stop their calculations 20 days after the protostar formation because of extremely small timesteps resulting from the high velocity of the jet propagating from the finest level. \cite{banerjee:06} also performed 3D collapse calculations down to stellar scales, using ideal MHD as well as molecular cooling, and they confirmed the results of \cite{tomisaka:02} and  \cite{machida:06}.

All the multidimensional studies we mentioned so far were done assuming a piecewise polytropic equation of state to mimic the thermal behaviour of the gas throughout the evolutionary sequence.  Such barotropic equation of states are parameterized using the results of 1D spherical symmetry simulations previously mentioned, so that they cannot account for multidimensional effects such as the optical depth drop in the vertical direction that allows efficient cooling of the nascent protostellar disks and thus leads to incorrect results for the fragmentation  \cite[e.g.,][]{boss:00,commercon:10,tomida:10}. 

The next level of complexity is to add a more accurate model for the thermal balance of the gas and dust mixture, i.e., to model radiative transfer (see Sec.~\ref{Sec:rad_trans}). The addition of a more accurate model for radiation transport can be crudely resumed to an additional heating/cooling term in the gas internal energy equation due to matter/radiation interactions. For the models which are restricted to the first collapse and first core formation, i.e., for temperatures less than $\simeq 2000$~K, the addition is conceptually straightforward for the gas evolution equations. An usual ideal gas equation of state can be used, with the only modification on the adiabatic index $\gamma$ which varies with the temperature. At low temperature ($T\lesssim 150$~K), the H$_2$ molecule behaves like a monoatomic gas ($\gamma=5/3$) because only the vibrational degrees of freedom are excited. At higher temperature, the rotational degrees start to be excited and the adiabatic index decreases to $\gamma=7/5$. Accounting for this non-constant adiabatic index is important regarding the fragmentation properties \citep{commercon:10} and the first core properties \citep{tomida:10,lee:18}. The driver of the second collapse is the endothermic dissociation of the H$_2$ molecule, which modifies considerably the gas equation of state. 

Currently, there are two ways to integrate non-ideal EOS in second collapse calculations. The first one is to compute on-the-fly the thermodynamic quantities from equilibrium chemical abundances and to use an ideal gas equation of state, including the effects of dissociation and ionisation, to compute the gas pressure and the internal energy \citep{black:75}. Most of the  second-collapse including RHD are done with this approximation \citep{masunaga:00,whitehouse:06,stamatellos:07,forgan:09,tomida:13,bhandare:18}. The second one consists in using tabulated EOS table coming from detailed studies to account for non-ideal effects (e.g., ionisation by pressure, interaction between particles) which are important at high density/pressure \citep{saumon:95}. \cite{saumon:95} EOS is used in the series of papers by \cite{vaytet:13,vaytet:14,vaytet:17,vaytet:18}, and also in the SPH public code PHANTOM \citep{price:17}. 

We note that these developments imply that the dependency of the specific heat capacity as a function of the physical conditions has to be taken into account. Caution should thus be taken within the  radiation solvers which are often integrated using implicit schemes. The most common and simple way to deal with it is to compute a heat capacity-like factor, $\tilde{C_\mathrm{V}}\equiv e/T$ where $e$ is the gas internal energy density, which is kept constant in the radiation solve. Readers can refer to \cite{tomida:13} for a detailed implementation. Unfortunately, details on the numerical implementations are usually not reported in the literature. 

The first full 3D RHD calculations of the second collapse of an initial 1~\msun~ dense core were performed by \cite{whitehouse:06} using the FLD approximation in a SPH code, followed by \cite{stamatellos:07} who used a local radiative cooling approximation and SPH. \cite{bate:10} extended the work of \cite{whitehouse:06} beyond the formation of the stellar core for about 50 yr, using $3\times10^6$ equal-mass SPH particles to satisfy the \cite{bate:97} mass resolution criterion throughout the entire collapse. Interestingly, there is to date no 3D RHD calculations of the second collapse, i.e., without magnetic field, performed with a grid-based code.

Last but not least, it has been demonstrated in \cite{commercon:10} that the interplay between magnetic fields and radiative transfer is of primary importance for the thermal budget of the collapsing gas and of the first and second core accretion shocks \citep{commercon:11,vaytet:12,vaytet:13,vaytet:18}. The infall velocity is indeed greatly modified in the presence of magnetic fields which tends to focus the collapsing gas on a smaller area compared to the case without magnetic fields \citep{commercon:11a,myers:13}. The magnetic braking mechanism transports angular momentum from the inner parts of the collapsing cloud to the envelop. As a consequence, the infall velocity increases along with the incident kinetic energy, which then modifies the thermal budget through the accretion shocks. The next step in the increasing complexity is to perform full 3D RMHD calculations. The first SPH RMHD calculations of protostellar core formation were performed in \cite{bate:13}, extending their previous work to ideal MHD.  \cite{tomida:13,tomida:15} performed the first 3D full RMHD calculations using a grid-based code, extending the work of \cite{machida:06} to account for radiative transfer with the FLD approximation, as well as the Ohmic dissipation and the ambipolar diffusion. They used a finite-volume nested-grid code with standard MUSCL predictor-corrector scheme to achieve second-order accuracy.

More details on \cite{tomida:15} implementation for non-ideal MHD are given in Sec.~\ref{Sec:nimhd}. For the implicit radiation update, they used a linear system solver with a combination of the BiCGStab solver and the incomplete LU decomposition preconditionner. \cite{tomida:13} used a nested-grid constructed to ensure that the Jeans length is always resolved by at least 16 cells, with typical spatial resolution at the maximum level of refinement of $\Delta x\simeq 6.6\times10^{-5}$~AU (23 levels). When this resolution condition cannot be satisfied, they stopped the calculations. This is particularly the case when discs form and grow with time to spread over different level of refinements. As a consequence, they cannot follow the second collapse in their model with ambipolar diffusion since a large disc is formed. More flexible refinement techniques, such as AMR or SPH are thus required to continue the calculations with a reasonable numerical resolution. Currently, the state-of-the-art includes 3D RMHD calculations with resistive effects, done with both SPH \citep{tsukamoto:15,wurster:18} and AMR \citep{vaytet:18}. The typical resolution in SPH calculations currently achievable with limited CPU resources is of $3\times10^{6}$ equal-mass particles per \msun. In AMR, \cite{vaytet:18} used a coarse grid of $64^3$ with 21 additional levels of refinement, and a refinement criterion of 32 points per Jeans length. While the SPH calculations are able to integrate the three non-ideal effects \citep{wurster:18} down to stellar scales, there is no AMR work describing the Hall effect in second collapse calculations (\cite{vaytet:18} accounts for the ambipolar diffusion and the Ohmic diffusion). 
The calculations by \cite{wurster:18} were integrated 17 yr after the stellar core formation, while \cite{vaytet:18} could only integrate for 24 days. Beside the major achievement which represents the formation of a protostar on a computer, we must admit that these kind of studies are not yet capable to give long term predictability. In addition, the CPU time required to integrate full 3D RMHD models is relatively big. 

Clearly, second collapse study cannot be done on a frame covering a large parameter space. In addition, even at the first core scale, the integration timestep becomes so short in the adiabatic fragments that long time evolution calculations becomes prohibitive. In that view, the sink particles are a good alternative to reach a long horizon of predictability and to cover a wide parameter space.

\section{Sink particles}

 Sink particles are Lagrangian particles, and can accrete matter as well as angular momentum in order to ensure mass and angular momentum conservation. Sink particles should not be confounded with sink cells which are usually introduced in grid-based codes which use a spherical grid to deal with the singularity in the center  \citep[e.g.,][]{boss:82,kuiper:10,li:11}. The differences between sink particles and sink cells are discussed in Sec.\ref{Sec:sink_discussion}. Sink particles are often erroneously referred as protostars or stars in the literature whereas the resolution at which they are introduced is generally much larger than the stellar scales, typically $1-10$~AU for collapse calculations, and a few $100$~AU for cluster formation and ISM evolution calculations. Multiple systems (small clusters, binaries), as well as  systems made of a disk around a star can thus be encompassed within a single sink particle depending on the resolution. 

\subsection{Standard implementations}

Sink particles were introduced in star formation calculations by \cite{bate:95} to enable long time integration to study the fragmentation of collapsing clouds. Sink particles were introduced to mimic the second collapse and the protostars at scales that could not be reached in numerical simulations. They are treated as accreting non-collisional point masses and enable one to follow the evolution through the main accretion phase, with a view to compare with observations of clusters in which most of the gas has been accreted. Their original implementation was done in a SPH code, with three main parts in the algorithm: i) sink particle creation, ii) sink particle accretion, iii) boundary conditions for sink particles. A SPH gaseous particle is tagged for sink particle creation if its density exceeds a specific density threshold (defined by the user) and satisfied a number of tests. First, the smoothing length of the SPH particle that is a candidate for sink particle creation has to be less than half the accretion radius $r_\mathrm{acc}$ of the sink particle. This ensures that the sink particle is formed from at least $N_\mathrm{neigh}$. $r_\mathrm{acc}$ is defined before the calculations start and remains fixed. Its value is chosen by the user depending on the required level of resolution. It sets the smallest scale that can be resolved in the calculations. The knowledge of the flow within $r_\mathrm{acc}$ is lost and the gas is assumed to collapse beyond this scale to form protostars. Then, a series of tests is performed on the system composed of the particle and its neighbours to decide if it should create a sink particle. The tests ensure that the gas particles would continue to collapse if the sink particle was not created. First, the ratios $\alpha$ and $\beta$ of the thermal energy and rotational energy to the gravitational energy must satisfy
\begin{equation}
\alpha\leq 1/2~;~ 
\alpha+\beta\leq 1.
\end{equation}
Additionally, the total energy must be negative, and finally, the system must be contracting ($\nabla \cdot \vec{v}<0$). If all the tests are passed, a sink particle is formed from the system at the centre of mass. Initially, the sink particle has thus a mass of $\sim N_\mathrm{neigh}$ times the standard SPH particle mass. Afterwards, sink particles interact only with the gas particles via gravity. The sink particle is  allowed to accrete mass at each integration timestep if a gas particle enters $r_\mathrm{acc}$ and passes several criteria: the particle must be the most bound to the sink particle (and not to another more distant but more massive sink particle), and the particle must have a moderate specific angular momentum.  When a gas particle is accreted, its mass and linear momentum are added to the sink particle, as well as the angular momentum which modifies the spin of the sink particle. The sink particle is moved to the center of mass of the sink and the accreted gas particle. Sink particles also contribute to the computation of the total gravitational potential in the same way as standard gaseous particles. 
\cite{bate:95} pointed out that the introduction of sink particles in SPH calculations may affect the gas outside the accretion radius, in particular because there is a discontinuity in the number of particles across the accretion radius. They proposed different types of boundary conditions to account for missing neighbours. Note that the use of boundary conditions in modern sink particles implementation is not used anymore thanks to the improvements that have been made, in particular in the case of optically thick flows  \citep[e.g.,][]{hubber:13}. 

Sink particles were introduced in Eulerian calculations for star formation purposes by \cite{krumholz:04} in the AMR code ORION. They introduce a Jeans length criterion for sink creation, based on the resolution study of \cite{1997ApJ...489L.179T}. A sink is created within a cell at the maximum level of refinement if the density $\rho$ exceeds the maximum Jeans density $\rho_\mathrm{J}$ that can be resolved
\begin{equation}
\rho_\mathrm{J}<J^2\frac{\pi c_\mathrm{s}^2}{G\Delta x^2_\mathrm{min}},
\end{equation}
where $J<0.25$ is the Jeans length resolution criterion, $c_\mathrm{s}$ the gas sound speed, and $\Delta x_\mathrm{min}$ the minimum cell size. The initial mass of the sink is $m_\mathrm{sink}=(\rho-\rho_\mathrm{J})\Delta x_\mathrm{min}^3$. In \cite{krumholz:04}, no additional checks are performed to validate the sink creation. Their algorithm is coupled to a sink merging scheme based on a friends-of-friends (FOF) algorithm to handle situations where a block of contiguous cells create sink particles in a single time step. The FOF is performed after each time step to group all the sink particles (old and new), with a linking length equal to $r_\mathrm{acc}$. All the groups of sink particles found by the FOF are then merged and replaced by a single particle at the center of mass of the group. All the merged sink particles quantities (mass, momentum, angular momentum)  are also added conservatively to the new sink particle. \cite{krumholz:04} sink accretion scheme was designed to handle situations where the flow onto the sink particle is subsonic (which is analogous to the case where boundary conditions have to be introduced in standard SPH implementation). They set the accretion rate using an approximate formula from Bondi-Hoyle accretion which handles also the regime where the flow is supersonic. The accretion rate is estimated using average properties in the region within $r_\mathrm{acc}$. An important point in \cite{krumholz:04} implementation is the description of the accretion zone. The mass accreted by the sink particle comes from all the cells in this accretion zone. Based on their experiments, \cite{krumholz:04} use a value of $r_\mathrm{acc}=4\Delta x$. In the case where the Bondi-Hoyle radius is smaller than the accretion zone, it is incorrect to estimate the accretion rate from all the cells in the entire accretion zone. \cite{krumholz:04} thus use an accretion kernel where each cell within $r_\mathrm{acc}$ is assigned a weight from a Gaussian-like function. The kernel is used to compute the average quantities to determine the mass accretion rate. The mass to be accreted is redistributed within the accretion zone onto virtual cloud particles (each cell in the accretion zone is divided into $8^3$ cloud particles).  Additional checks can be then performed to avoid to accrete gas that has for instance too much angular momentum to be gravitationally bound. When a parcel of gas is accreted, it also transfers its linear and angular momentum to the sink. An absolute mass cap is also used to limit to 25\% the amount of mass accreted from a unique cell in a single time step.  

After accretion, the position of the sink particle is changed in two steps according to its momentum after gas accretion and through gravity. First the momentum is updated to account for the gravitational interaction with the gas. The total force on a sink particle is computed by a direct summation over all the cells in the computational domain, except the ones in the accretion zone. This method remains practicable as long as the number of sink particles created remains limited. For the accretion zone, the interaction is computed between the sink and the cloud particles created during the accretion step. A Plummer law  is used to soften the gravitational interaction at short distance, i.e., $F_\mathrm{grav} \sim 1/(r^2+\epsilon^2)$ where $\epsilon=2\Delta x_\mathrm{min}$ is the softening length. Second, the acceleration due to particle-particle interaction is added to the sink particle momentum using a Bulirsch-Stoer algorithm with adaptive time step and error control.
Note that the stability of the temporal discretisation scheme required that the sink particle velocity satisfies a CFL-like condition ($v_\mathrm{sink}\Delta t < C\Delta x$, with $C<1$). 

Sink particles algorithms have been complemented in the past ten years to improve their versatility, in particular in the transition between the sub- and super-sonic regimes. All newer implementations have been built upon these two pioneering works.

\subsection{Contemporary implementations}

In all current implementations, the sink creation and accretion are the most sensitive algorithms. Of course, the sink particles can be introduced in the initial conditions. Otherwise, they are created after the creation checks. 

Each of the commonly used codes in star formation has its own implementation of sink particles. For the grid-based codes, we note the work of \cite{wang:10} in the ENZO code, which is very similar to \cite{krumholz:04} but with a different merging scheme which involves a maximum mass sink mass for merging.  \cite{federrath:10} presented an implementation in the AMR code FLASH, where the creation checks were similar to the ones by \cite{bate:95} for SPH with two additional checks: the cell tagged for sink creation should be at the maximum level of refinement, and it should sit in a local gravitational potential minimum. They also account for magnetic energy in the energy budget done in the check algorithm. The checks are performed on a spherical region of radius $r_\mathrm{acc}$ around the cell candidate for sink creation. Their scheme for accretion is also different from the previous grid-based works. If a cell $i$ within $r_\mathrm{acc}$ exceeds the density threshold $\rho_\mathrm{res}$ for sink creation, the mass increment $\Delta M=(\rho_i-\rho_\mathrm{res})\Delta x_i^3$  is considered for accretion onto the sink particle. The mass increment has to be gravitationally bound to the sink, and in the case of overlapping sink particles, it is accreted to the most gravitationally bound sink particle, the latter being moved to the center of mass of the particle-gas configuration before the accretion step. They also improve the treatment of the gas-sink gravitational interaction. Similarly to \cite{krumholz:04}, the sink-sink interaction is done by direct summation over all the sink particles. The gravitational acceleration of the sink due to the gas is estimated from the gravitational potential of the gas which is handled by the Poisson solver of FLASH to handle the gas-gas interaction. The acceleration is interpolated using a first-order cloud-in-cell method for each sink particle. Last, the acceleration of the gas due to the sink is done by a direct summation. Each sink particle contributes to the acceleration of each cell in the computational domain. The direct summations for the sink-sink and sink-gas interactions requires a gravitational softening. \cite{federrath:10} use a cubic spline softening similar to what is used in SPH, which is less aggressive within the softening length than a simple Plummer profile. \cite{federrath:11} updated their first implementation to ensure exact momentum conservation using a direct summation between all cells and sink particles for the gas-sink interaction as well. 

In addition, \cite{federrath:10} performed a quantitative comparison between their implementation and the one by \cite{bate:95} by comparing the results of two star formation calculations performed with FLASH and with a SPH code. They find a good quantitative agreement which was quite encouraging given the fundamental differences between SPH and AMR sink implementations.

\cite{hubber:13} presented an improved algorithm for the sink particles in SPH, which can handle situations where sinks are
introduced after the gas has become adiabatic. They propose a different combination of creation criteria which are based on the studies we just mentioned: i) density threshold, ii) no overlapping with another sink accretion volume, iii) the gaseous particle flagged for sink creation sits in a minimum of gravitational potential (on a volume defined by its neighbours), iv) the gaseous particle is outside the Hill sphere of fragments  harbouring another already formed sink. 

Last, \cite{bleuler:14} presented an innovative sink creation scheme based on a clump finder algorithm used on-the-fly in the RAMSES code.
The clump finder is performed in a first step to identify peaks and their associated regions.  Second, the peaks are considered for sink creation. A virial theorem type analysis is performed on the clumps, accounting for surface pressure as well as tidal forces.  The virial check is passed if the second derivative of the moment of inertia in the center of mass frame is negative. Then a collapse check is performed, which corresponds to the contraction check of \cite{krumholz:04} and \cite{federrath:10}, but adapted according to the virial analysis. Last, the final check is a proximity check. 
We also note that \cite{bleuler:14} offer also the possibility to use the alternative sink creation and accretion schemes reported in the literature.  They provided an excellent review of the different creation checks, accretion, sink merging and trajectories schemes. In particular, they tested different schemes to perform sink accretion: Bondi-Hoyle accretion, flux accretion, and density threshold accretion.  They show that all methods give good results for spherical Bondi accretion (i.e., when the flow is supersonic). In the subsonic regime, only the Bondi formula gives a correct accretion rate. They recommend to automatically adapt the accretion scheme depending on the surrounding flow properties. 

A number of other sink particle implementations have been reported in the literature that we cannot detail in this review. We give in the following a non-exhaustive list, restricted to the field of star formation. In SPH, sink particles have been implemented in the codes GADGET  \citep{jappsen:05}, DRAGON \citep{goodwin:04}, and SEREN \citep{hubber:11}. For grid-based codes, sink particle are used in the uniform grid code ATHENA  by \cite{gong:13}, in RAMSES by \cite{padoan:11}, and in the ORION2 code by \cite{lee:14}. Last, \cite{greif:11} also proposed a sink particle algorithm within the AREPO moving-mesh code. Their implementation is very similar to the one used in SPH codes. We also note that sink particle merging is now widely used in most of the implementations. 

Twenty years after their introduction, sink particles remain a sensitive subject in the community. It is commonly accepted that an accreting sink particle should have a minimal impact on the collapsing gas around it. The easiest way to deal with it is to create the sink particle during the (quasi-) isothermal phases of the collapse (either the first or the second collapse), when the gas velocity is supersonic. The perturbation created by the sink particles on the gas dynamics cannot propagate since the CFL timestep is limited by the supersonic motions. An example of bad behaviours observed in protostellar collapse SPH calculations using sink particles introduced in the vicinity of the first cores can be found in Appendix D of \cite{commercon:08}. When a barotropic law is used to mimic the gas thermal budget, a simple method used to introduce sink particles in the vicinity of hydrostatic object can be found in \cite{price:08}. They artificially change the gas equation of state to isothermal at an arbitrary density (e.g. $10^{-11}$~g~cm$^{-3}$) and set the density threshold for sink creation to a value of more than two orders of magnitudes larger.

\subsection{Sink particles and magnetic fields}

In the previous section, we did not discuss the sink particle algorithms in the presence of magnetic fields. We try to describe here some attempts done to provide a good description of MHD flows for sink creation and accretion. Sink particles can of course be introduced in MHD calculations, but the implementation is not as straightforward as in the hydrodynamical case. While it is relatively easy to accrete gas and momentum, magnetic flux accretion cannot bypass the solenoidal constraint $\nabla \cdot \vec{B}=0$. 

\cite{price:08} and \cite{wang:10} performed star formation calculations in turbulent and magnetized clouds, but do not account for any contribution of the magnetic fields in their sink algorithms. \cite{federrath:10} have introduced the magnetic energy in their negative total energy check for sink creation, but not in the accretion scheme. 
\cite{lee:14} have extended the work of \cite{krumholz:04} to deal with magnetized flows. For sink creation, they follow the work of \cite{myers:13} who derived a magnetic Truelove criterion. They derive simple steady state accretion rates as a function of the plasma beta $\beta$ and the Mach number $\mathcal{M}$. 

In all implementations, gas is accreted by the sink particles but not magnetic flux. It stays the gas in the surrounding of the sink particles, and magnetic flux accumulates as collapse goes on.  While at scales of a few 10-100s AU this flux accumulation is not problematic \citep{federrath:12,li:18}, it eventually leads  to a magnetic explosion and magnetic flux redistribution in the surrounding of the sink \citep{zhao:11} due to the development of an interchange instability \citep{spruit:90,spruit:95,li:96}, which is also observed in high resolution models without sink particles  and ideal MHD \citep{li:14,masson:16}. In addition, since density stays roughly constant while magnetic intensity increases in the sink accretion zone, the Alfv\'en speed increases significantly, leading to non-physical accretion rates, as well as short integration time steps. \cite{lee:14} proposes to cap the accretion rate at a maximum value corresponding to a constant Alfv\'en speed after mass accretion. This trick tempers the accumulation of magnetic flux as well. 

In the recent years, resistive MHD calculations have shown that all non-ideal effects leads to a decrease of the magnetic flux at the first core scale. Magnetic fields decouple from the gas and dust mixture dynamics. The magnetic flux does not accumulate at the first core border, but rather at a larger distance which corresponds to the radius at which non-ideal effects start to dominate \citep{masson:16, hennebelle:16}. Matter is accreted but leaves its associated magnetic flux outside the decoupling region. Adding sink particles in restive MHD calculations with resolution capable of resolving the decoupling region has not yet been investigated in detail \citep{machida:09,matsumoto:17,tomida:17}, but one can foresee that introducing a sink within the decoupling region should  prevent magnetic flux accumulation. It is expected that non-ideal MHD can alleviate the conditions of interchange instability, but is is then necessary to resolve the decoupling scale, i.e., the disk. To date, the most detailed study of sink particles introduction in MHD calculations can be found in \cite{machida:14}.

\subsection{Sink particles versus sink cells}
\label{Sec:sink_discussion}

A fundamental difference between sink cells and sink particles is that sink cells are fixed in position and have physical boundaries. There are pros and cons for the two methods, which we briefly summarize in the following.

First, the sink cell can be considered as a boundary in codes using a spherical grid.  In collapse calculations, a sink cell is most often introduced as a passive boundary with zero gradients, or also called outflowing boundary conditions \citep{yorke:02,kuiper:10,li:11}. The mass flux across the boundary is added to the central point mass of the sink cell and used to compute the gravitational acceleration. The sink cell enables one to make more detailed models for the magnetic field configuration at the boundary, as for instance in star-disk interaction studies, \citep[e.g.][]{mellon:08,takasao:18}. In principle, this is a quite powerful numerical set-up to overstep the issue of time integration after the second core formation. The current state-of-the-art is not yet mature to design such numerical studies, but this should certainly be the purpose of future works.

On the other hand, a major inconvenience of sink cells resides in their fixed position in time, which forces the center of mass to be equal to the geometrical center of the disk. As a consequence, the development of non-axisymmetric azimuthal low wave-number modes may be artificially damped, which alters the development of eccentric gravitational instabilities \citep{adams:89, shu:90}. This effect can be particularly  problematic in the case of young stellar objects with disks of mass comparable to the stellar mass, $M_\mathrm{disk}/M_\mathrm{star}>1/3$ \citep{krumholz:07,kuiper:11,sigalotti:18}.

Last, the use of "outflow" boundary conditions for the mass accretion on the sink cell is very different from the algorithms we just presented for the sink particles.  \cite{machida:14} have shown that applying similar criteria for the sink accretion than the ones employed in studies with a sink cell \citep{li:11} may lead to very different results, i.e., no disk formation, compared to the same calculations done with a classical sink particle algorithm, i.e., formation of a disk.

\section{Sub-grid models for sink particles}
\label{Sec:sink_subgrid}

Beyond the practical advantage of saving computational time, the sink particle has been shown to be very useful in works where (proto-)stellar feedback is considered. Similar to cosmological and galactic evolution studies, the sub-grid models can be associated to the sink particle evolution. We distinguish two types of sub-grid physics, depending on the feedback origin: radiative (luminosity, accretion shock, ionisation), and dynamical (jets/outflow, winds). In any case, the source terms introduced by sub-grid physics are treated using operator splitting schemes.

\subsection{Radiative feedback}
The first class of sub-grid models were designed to account for the radiative feedback of the nascent protostars. The radiative feedback term includes the accretion, the internal, and the ionisation luminosities. When integrating over discrete timesteps, the luminosity is an energy or radiative flux input. Radiative feedback was introduced early in 2D models \cite[e.g.,][]{bodenheimer:90}. They used a 2D axisymmetric hydrodynamical code \citep{rozyczka:85}, which was primarily designed for stellar winds. \cite{bodenheimer:90} used the grey FLD approximation for the radiation transport. They combined the evolution of their central zone, or sink cell, with a pre-main sequence evolution model based on Maclaurin spheroids in hydrostatic equilibrium. They computed the structure of the protostar according to the central $(\rho, T)$ path tracks reported by \cite{winkler:80} from 1D spherical calculations. The protostar mass $M_\star$ and equatorial radius $R_\star$ were then used to compute the accretion luminosity $L_\mathrm{acc}=0.5 GM_\star \dot{M}/R_\star$, where $\dot{M}$ is the mass flux crossing the inner boundary. \cite{bodenheimer:90} further assumed that the radiative energy input $L\Delta t$ was only due to the accretion luminosity and that all the infall kinetic energy was converted into radiation at the stellar core accretion shock. They found that most of the energy that went for heating in the envelope comes from this protostellar luminosity. 

\cite{yorke:99} and \cite{yorke:02} used a more sophisticated protostellar evolution model which accounts for the intrinsic luminosity of the protostar $L_\mathrm{int}$. They realised that the central luminosity evolution sets the thermal budget within the envelop, but could not resolve  the innermost regions of the star-disk system. They designed a protostellar accretion luminosity model which accounts for the luminosity coming from Keplerian motion of the disk being converted into heat and radiated away at the disk-core boundary layer following \citep{adams:86}
\begin{equation}
L_\mathrm{tot}=L_\mathrm{int} + \frac{3GM_\star \dot{M}}{4 R_\star},
\end{equation}
where they assume that a fourth of the total potential energy of the accreted material is released into heat within the disk, the remaining being radiated away within the unresolved inner disk region and at the accretion shock. In addition, \cite{yorke:99} and \cite{yorke:02} followed radiation transport using a frequency dependent ray-tracing algorithm (with 64 frequency bins). In particular, \cite{yorke:02} demonstrated the importance to handle the protostellar irradiation using a frequency dependent irradiation scheme in the context of massive star formation. The flashlight effect, i.e., radiation escapes in the vertical direction and the radiative acceleration is reduced within the disk to allow accretion, is enhanced compared to a simple grey model. In addition, \cite{krumholz:05} showed how protostellar outflows allow radiation to escape in the outflow cavities to allow a continued accretion onto the central massive stars. These results have been further confirmed by 3D dynamical simulations of several authors \citep{krumholz:07,kuiper:11,rosen:16,klassen:16,kuiper:16}.

Currently, most 3D numerical RHD calculations include stellar radiative feedback sub-grid models attached to sink particles/cells evolution. Most of the implementations account for both the internal and the accretion luminosities, with some modulation factors on the amount of potential energy radiated away and on the efficiency of the mass transfer for the sink accretion radius to the protostar. The simpler sub-grid models use a grey FLD model for the radiation transport where the protostellar luminosity is simply a source term in the radiative energy equation \citep{krumholz:07,offner:09,stamatellos:11,fontani:18, jones:18}. Frequency dependent protostellar irradiation modules have been developed using ray-tracing to compute the radiation fields, combined with moment models for radiation hydrodynamics \citep{kuiper:10,2011MNRAS.414.3458W,klassen:14,ramsey:15,buntemeyer:16,rosen:17}.  These  frequency dependent modules have been used primarily  in the context of massive star formation \citep{rosen:16,klassen:16}. 

Last, we note the developments made to capture ionising radiation feedback from massive stars: \cite{dale:07} used a Str\"omgren volume method in SPH, \cite{peters:10}  ray tracing in the FLASH code, \cite{kuiper:18} an hybrid ray-tracing and FLD irradiation module in the PLUTO code, \cite{geen:15} the M1 moment method in the RAMSES code, and \cite{harries:15} Monte Carlo radiative transfer in the AMR code TORUS. These schemes are applied first in isolated collapse calculations \citep{kuiper:18} and, mostly, in cluster formation studies \citep{peters:11,dale:11,dale:12,geen:15,2017MNRAS.472.4155G,ali:18,geen:18}.

\subsection{Dynamical feedback}

The most advanced MHD collapse calculations have shown that outflows and jets naturally develop at different scales during the collapse \citep{tomisaka:02,banerjee:06,machida:06,hennebelle:08,ciardi:10,commercon:10,tomida:10,price:12}. In addition, outflows and jets are commonly observed in YSO's (it is for instance a selection criterion for Class 0 sources, \cite{andre:93}) and they are considered as possible sources of turbulence driving in star forming clouds and participate in the regulation of the star formation rate \citep{matzner:07,nakamura:07,krumholz:14,federrath:15}. Outflows and jets launching scales are nevertheless not captured in most studies because it requires a very high numerical resolution to reach sub-AU scales.  Besides, studies neglecting magnetic fields cannot generate MHD winds driven centrifugally or by toroidal magnetic pressure. Consequently, very few numerical calculations have been able to launch self-consistently MHD outflows \cite[e.g.,][]{hennebelle:11}. Sub-grid models have thus been developed to account for outflows/jets  in numerical works with unresolved  launching scales and missing physics, similarly to what has just been presented for the radiative feedback. The numerical implementation are based on analytical works dedicated to MHD wind launching \citep{blandford:82,pelletier:92,shu:94,ferreira:97,matzner:99}. 

The first works on sub-grid models have been reported in \cite{li:06} and \cite{nakamura:07} who employed ideal MHD and self-gravity. They attached a protostellar outflow model on sink particles following \cite{matzner:00} because they were unable to describe the self-consistent launch of outflows with the crude resolution they use ($128^3$ for a $\approx2$~pc box). Each star injects in the ambient medium a momentum that is proportional to the stellar mass $M_\star$ with a two-component outflow model. The volume of injection is given by the sink accretion volume (which they refer to as "supercell"). A fraction $\eta$ of the total outflow momentum is put in a conical collimated jet component to facilitate the transport of energy and momentum at large distances. The rest of the momentum is put into a spherical component around the protostar. The direction of the jet is given by the orientation of the magnetic field in the central cell, and they assume an opening angle of 30$^\circ$ about this direction. 

Most of the implementation of outflows and jets have a similar construction, with flavors depending on the sink particle algorithm. \cite{wang:10} implemented in the MHD ENZO code a simplified version for 
protostellar outflows where they only account for the collimated jet component. They adopt a continuous momentum injection $\Delta P=P_\star\Delta M$ where $P_\star$ is a proportionality constant which 
depends on the stellar mass $\sim M_\star^{1/2}$). If the momentum injection occurs in a cell with a too low density, they take 10\% of $\Delta M$ out of the sink particle and redistribute it evenly 
between the injection cells in order to avoid too large outflow speeds (and too small integration timestep). Along the same line, \cite{cunningham:11,hansen:12} extended the work of 
\cite{krumholz:05,krumholz:07} and presented for the first time a combined sub-grid model which includes protostellar outflows \citep{matzner:99} and protostellar radiative heating \citep{offner:09}, 
ignoring magnetic fields though. Their sub-grid model accounts for pre-main sequence evolution as in \cite{offner:09} and the calculated protostellar radius is used in the outflow and radiative 
heating models. They parameterized their outflow model with dimensionless parameters which, for instance, sets the fraction of mass accreted by the protostar or launched in the outflow, and the 
ejection velocity as a function of the Kepler speed at the surface. Contrary to the case of radiative heating where energy is put within the accretion volume, they inject outflows at a distance 
comprised between $4\Delta x$ and $8\Delta x$ from the sink particle,  and used angular dependency from \cite{matzner:99}. We note that they not only inject momentum, but also mass and thermal energy. 
They set the wind temperature to $10^4$~K, which is appropriate for an ionised wind. Using the same tool, \cite{myers:14} added magnetic field in their simulations with ORION, but because of the low 
resolution they used ($>20$~AU), they where not able to launch self-consistent outflows and used the sub-grid model we just mentioned. The outflows they produce may form strong shocks at very high temperature, 
higher than the dust sublimation temperature so that the dust opacity drops. Their radiative transfer scheme, primarily designed for dust thermal emission does not allow the gas to cool efficiently. 
They thus needed to add a line cooling function for the thermal budget of the gas.

\cite{federrath:14} implemented a sub-grid-scale outflow model upon the sink particles algorithm they implemented in FLASH \citep{federrath:10} for MHD collapse. Their model combines different features of the \cite{li:06,nakamura:07} and \cite{cunningham:11} implementations. They used a normalised velocity profile for momentum injection, which reproduces the two components outflow/jet with opening angles of $30^\circ$ and $5^\circ$ respectively. They also improved the algorithm for the outflow orientation by recording the angular momentum transfer from the accreted gas to the sink particle. Their algorithm also conserves mass and momentum exactly in the sink accretion step.
\cite{murray:18} proposed a similar implementation in RAMSES. They follow previous works \citep{cunningham:11,federrath:14,offner:09,myers:14} for the outflow injection as well as the protostellar evolution.  Outflows are there injected within a conical volume about the spin axis of the sink particle in a radial extent comprised between 4 and 8 cells (at the highest refinement level) away from the sink particle.  \cite{federrath:14} performed a parameter study on the number of cells for the radial outflow direction and found that convergence on the maximum outflow velocity requires 16 cells per outflow radius. Convergence on the mass, linear and angular momentum of the outflow is achieved though with 8 cells. 

All these protostellar outflow sub-grid models are currently widely used in the community for astrophysical applications on the star formation rate/efficiency \citep[e.g.][]{federrath:15,kuiper:16,offner:17,murray:18}, stellar initial mass function \citep{krumholz:12}, origin of turbulence driving \citep{nakamura:11,offner:18}, as well as cluster and massive star formation \citep{wang:10,cunningham:11,kuiper:15,li:18}.

In the context of massive star formation, another class of outflows models are based on the luminosity and ionisation. Sub-grids models of outflows generated by massive star luminosity have been developed \cite[e.g.][]{krumholz:05,peters:14}, but the physical mechanisms driving this type of outflows are not yet  well understood. No work in the context of massive star formation has accounted for the combined effects of magnetic fields and radiative force with a resolution sufficient to resolve the magnetic outflow launch. There is certainly a combination of magneto-centrifugal process and from radiative pressure, depending on the mass of the forming protostar.

\subsection{Second collapse and pre-main sequences}

Ideally, the sub-grid models should be designed to reproduce the results of high resolution studies, which resolve the protostar scales. The short time integration after stellar core formation, dominated by the adiabatic evolution, does not allow to design sub-grid models for protostellar feedback from full 3D RMHD calculations with a high level of confidence. For these reason, all recent works mostly rely on the results of 1D spherical symmetry results to set their  protostellar core evolution models. For instance, \cite{kuiper:13} use the pre-main sequence evolution code STELLAR \citep{bodenheimer:07,yorke:08} to determine the protostellar luminosity of the non-resolved protostars. \cite{pelupessy:13} developed the open source Astrophysical Multi-purpose Software Environment (AMUSE), a component library of simulations involving different physical domains and scales. For instance, it enables one to couple hydrodynamical codes (GADGET, ATHENA, AMRVAC) with pre-main sequence and binary evolution codes such as MESA \citep{paxton:11}. Recently, \cite{wall:19} used AMUSE to perform a star cluster formation model that includes individual star formation from self-gravitating, magnetized gas, coupled to collisional stellar dynamics. It couples different tools: the AMR code FLASH \citep{2000ApJS..131..273F}, the N-body code ph4 \citep{mcmillan:12}, and the stellar evolution code SeBa \citep{portegies:96}. We anticipate to see more works using this strategy of coupling tools and scales in the near future. 
 
\begin{figure}[t]
\begin{center}
\includegraphics[width=17cm]{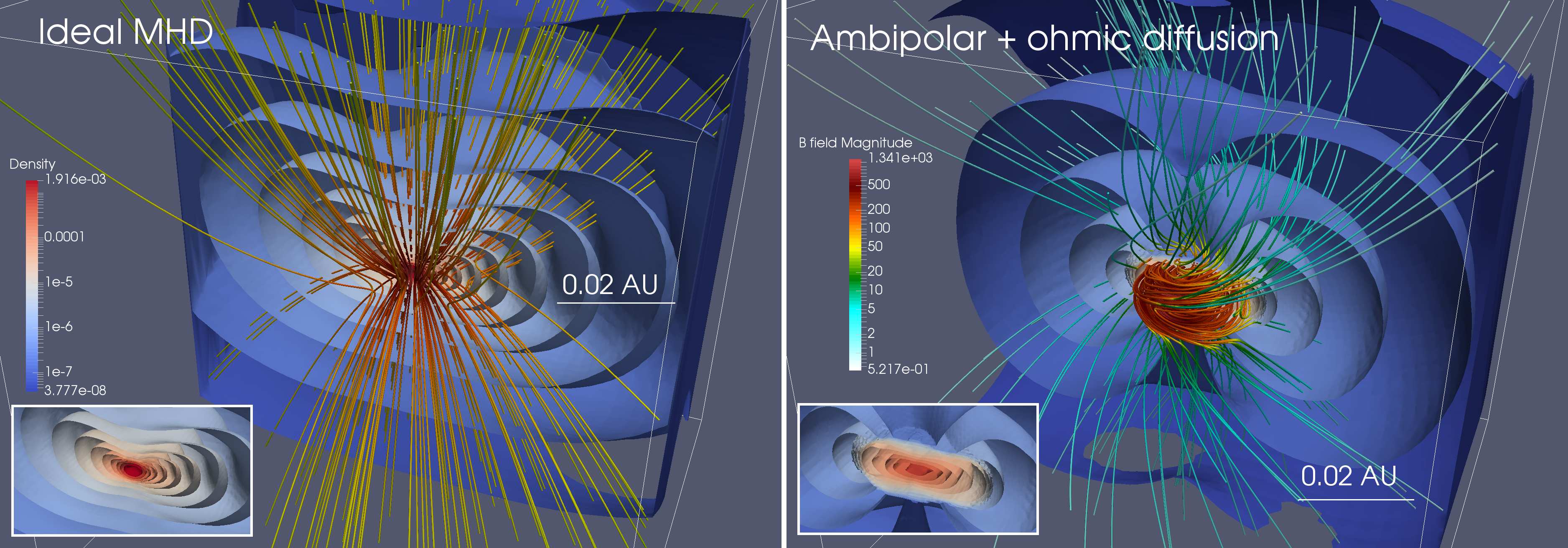}
\end{center}
\caption{Volume rendering from collapsing 1~\msun~dense cores at the protostellar core scale. The left panel shows the results of an ideal MHD 3D simulations, and the right panel the comparison with non-ideal MHD including ambipolar diffusion and Ohmic diffusion. The color lines indicate the 3D magnetic field lines, and the color coding the magnetic field amplitude.  Figure reproduced from \cite{vaytet:18} with permission of A\&A. \label{fig:vaytet}}
\end{figure}

Recent work from \cite{vaytet:18} have managed to resolve the second core formation with non-ideal MHD (Ohm diffusion and ambiploar diffusion) from collapse 1~\msun~dense core (see Figure~\ref{fig:vaytet}). Interestingly, they show that the magnetic field lines topology is very different at the first and second core scales. Ambipolar diffusion is dominating at the first core scale. The magnetic fields direction is essentially vertical and the magnetic fields and the gas evolution are decoupled. At the second core scale, the ionisation rises due to high temperature and the system is back to ideal MHD, but with a high plasma beta $\beta=P_\mathrm{therm}/P_\mathrm{mag}$. The weak magnetic field is efficiently wrapped up by the gas rotation and the resulting fields direction is essentially toroidal. On the opposite, when ideal MHD is preserved all the way from the dense core to the stellar core, the magnetic fields lines are strongly pinched and the field remains poloidal at the second core scale. The energy balance at the first and second core accretion shocks is also different. While the accretion shock at the first core has been classified as a radiative supercritical shock, with {\it all} the incident kinetic energy radiated away (or often referred to as cold accretion), in 1D spherical core collapse studies \citep{commercon:11,vaytet:12}), recent 3D works have shown that the first core accretion experiences both cold and hot accretion at the same time at its surface, depending on the accretion flows morphology. On the surface of the stellar core, the accretion shock has been found to be subcritical \citep{vaytet:12,tomida:13,vaytet:18} with a significant part of the incident kinetic energy transferred to the protostars. Although these results have been investigated only for the very first stages of the protostar evolution (only one month in \cite{vaytet:18}), they indicate that it is not easy to derive general properties of small scales phenomena to design robust sub-grid models. In particular, \cite{wurster:18c} show that the magnetic fields evolution of the protostars they formed in their SPH models is largely affected by numerical diffusion. Last, the radiative efficiency of the stellar accretion shock has been shown to be a key process for the pre-main sequence evolution and sets the radius of the young protostars  \citep{baraffe:12}. We recall that the sub-grids models designed for the radiative and dynamical feedback take the protostellar radius as an input. The launching speed of the protostellar jets as well as the accretion luminosity depends on $R_\star^{-1}$.

\section{Discussion}

We have described almost all the numerical methods that are required to model the star formation process within the turbulent interstellar medium.
These discretised equations must now be solved for a given set of initial and boundary conditions, over some prescribed time. This defines
the numerical set up and the overall simulation strategy. Obviously, given the wide range of spatial and temporal scales involved, one cannot 
realistically model an entire galaxy all the way down to the formation of brown dwarves and proto-planetary disks. Unfortunately, given the complexity of
the star formation process, and the fact that large and small scales are strongly coupled, this is required by the physics of the problem.
Indeed, supersonic turbulence is seeded on large scales, where kinetic energy is injected through shearing or colliding flows induced by galactic rotation, 
spiral waves and extra-galactic accretion flows. Moreover, turbulence and gas ejection can be triggered by stellar feedback processes occurring on very small scales,
such as radiation driven flows, jets and ultimately supernovae explosions. 

\begin{figure}[t]
	\begin{center}
		\includegraphics[width=17cm]{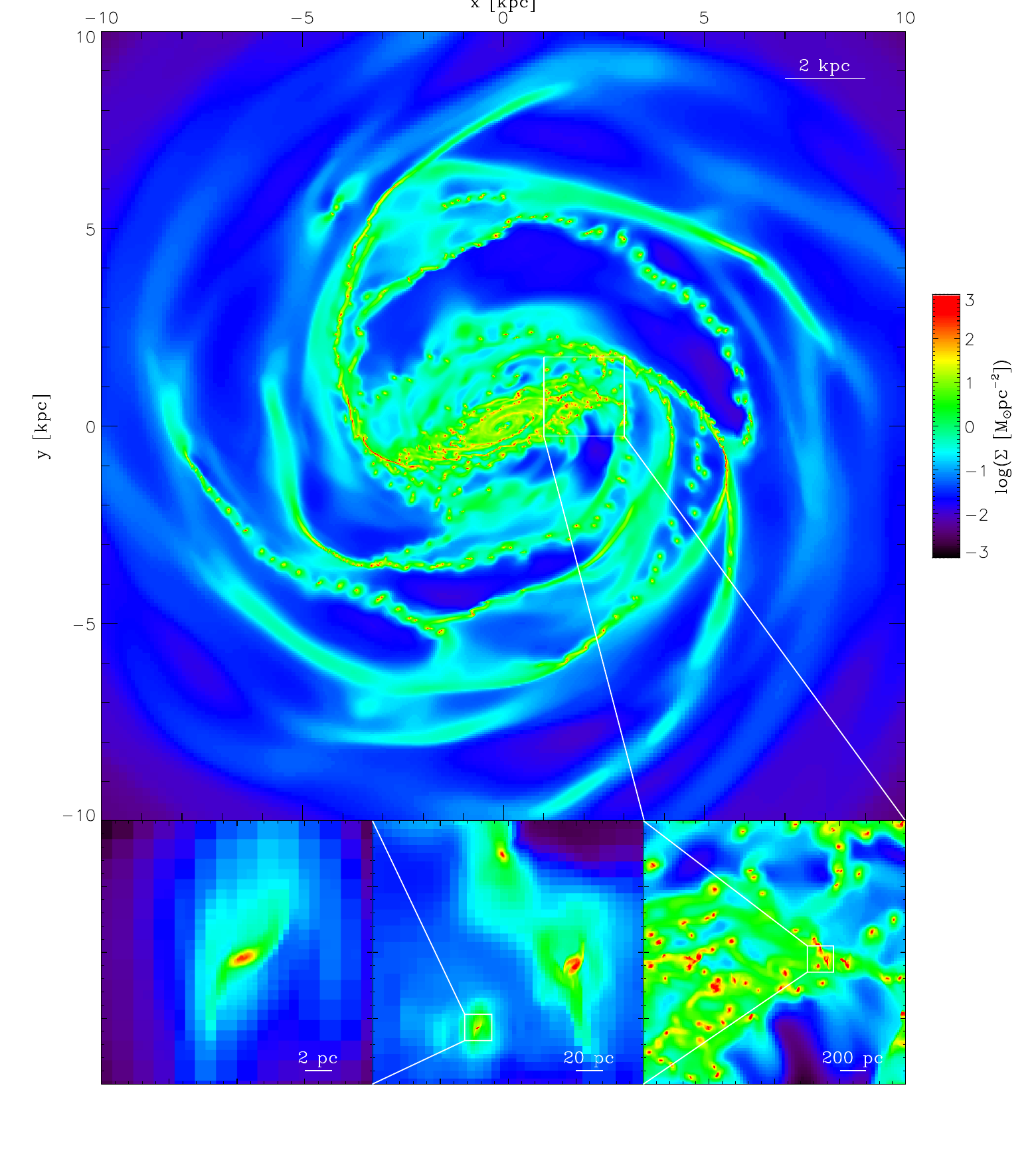}
	\end{center}
	\caption{Simulation of the Milky Way: column density of the gas disc in a sub-parsec resolution simulation. The colour table only applies to the main panel: the table has been changed in each zoom-in view to enhance contrast. Figure adapted from \cite{renaud:13} with permission from the authors. \label{fig:renaud}}
\end{figure}

Modelling star formation is therefore fundamentally a multi-scale, multi-physics problem, and a very difficult one. Past or present day simulations always rely on some compromises.
We can decompose them into several categories, corresponding roughly to the adopted box size and initial conditions. First, full galaxy simulations, where 
spiral waves and spiral shocks are used as the seed for turbulence. The box size must be of several tens of kiloparsecs to contain the entire
rotating disk and possibly the galactic corona. The gas is compressed through the spiral shocks or various colliding flows. Thermal 
and gravitational instabilities fragment the gas into small molecular clouds. The spatial resolution never really drops below 0.1 or 1 pc in these simulations,
so that stars cannot be modelled individually, at least for massive galaxies like the Milky Way (see Fig.~\ref{fig:renaud} for illustration). Simulators rely on simplified, sub-grid star formation recipes
based on  stochastically spawning star cluster particles of the same mass, or on the sink particle technique where in this case sink particles represent a star cluster,
rather than a single star. \cite{bournaud:10} studied the properties the ISM substructure and turbulence in galaxy simulations with resolutions up to 0.8 pc
and $5\times10^3$~\msun~with RAMSES. \citep{renaud:13,kraljic:14} were able to capture the transition from turbulence-supported to self-gravitating gas with resolution up to 0.05 pc  in simulations of Milky Way-like galaxies using RAMSES. 
\cite{hopkins:12} presented SPH simulations dwarfs galaxies and Milky Way (MW) analogues to massive star-forming galactic discs with pc-scale resolution. Later, \cite{hopkins:14} wrapped their implementation of feedback in the FIRE (Feedback In Realistic Environments) simulations suite, aiming, among others objectives, at  resolving the formation of giant molecular clouds  and the multi-phase interstellar medium (ISM).
More recently, \cite{hopkins:18} updated the FIRE implementation in the GIZMO code and performed sub-parsec resolution simulations of galactic discs. 
We also refer readers to the work of \cite{dobbs:18} using SPH, \cite{2011ApJ...730...11T} using ENZO and \cite{smith:18} using AREPO. In addition, we note that for dwarf galaxies, it becomes possible to model much smaller
scales so that at least massive stars can be modelled individually: \cite{hu:16,hu:19} using SPH and a resolution of $1-5$~\msun~by gas particle, \cite{emerick:18,emerick:19} using ENZO with a maximum resolution of 1.8~pc.

The second approach relies on simulating only a small portion of vertically stratified galactic disks. The box size is usually around 1 kilo-parsec, with periodic boundary conditions
in the direction of the disk plane and outflow conditions in the direction perpendicular to the disk plane. The geometry is clearly heavily constrained by this 
elongated, vertical and stratified layer of gas but it captures most of the phenomenon at work locally in the disk. Using the shearing box technique,
one can also add more realistic shearing conditions, so that turbulence can be maintained both by stellar feedback and a large scale galactic shearing flow \citep{kim:02}. 
In this case, the resolution can drop significantly below 0.1 parsec, may be 0.01 parsec or even less for the highest resolution simulations. 
Unfortunately, it is still impossible to model individual stars at this resolution so that here again most papers are based on the sink particle techniques for representing star clusters. 
It is however almost possible to resolve massive molecular cores (and their associated sink particles) individually, so that individual supernovae and HII regions can be resolved. These simulations are probably the most realistic
models of the interstellar medium of the Galaxy, although still far from resolving the entire stellar population. In standard setups, the models account for the thermal instability and feedback (mechanical by SN and radiative by massive stars). First grid-based models were presented in \cite{korpi:99} and included magnetic fields, SN heating and radiative cooling. 3D MHD AMR models of a kpc-scale ISM driven by SNe were presented in \cite{deavillez:05} and \cite{joung:06}. Currently, numerous studies and projects are based on a similar setup, with increasing physics put on over several years (SNe, radiative feedback, cosmic-rays, shear, etc...): the series of papers by \cite{hennebelle:14,iffrig:17,colling:18} as well as the work of \cite{martizzi:16} and \cite{butler:15,butler:17} using RAMSES, by \cite{kim:11,kim:13,kim:15,kim:17,kim:18} using ATHENA, the SILCC project papers using FLASH \citep{walch:15,girichidis:16,gatto:17,peters:17,girichidis:18}.

The third approach aims at simulating individual molecular cores. The box size ranges from 100 parsec down to 0.1 parsec for the highest resolution simulations. The mass typically ranges from 50 to a few $10^3$ \msun.  
Boundary conditions are either periodic or isolated. In the latter case, the cloud as a whole can collapse, while the former set of simulations represents a more uniform
background with collapsing regions only at very small scales. The smallest box sizes represents internal regions of larger clouds and are the only simulations for which 
the entire star population, down to the brown dwarf limit, can be modelled. These simulations are directly tackling the origin of the stellar IMF, and the formation 
of proto-planetary disks. We mention in the following a non-exhaustive list of the work done in this intense field of research. The series of papers by Bate \& collaborators have investigated  the collapse of molecular clouds of mass ranging from 50 to 500~\msun~with increasing physics and numerical resolution since their pioneer work \citep{bate:03} using SPH. In particular, they have investigated the effect of radiative transfer \citep{bate:09,bate:12} and magnetic fields \citep{price:08,price:09}. The work of \cite{dale:07,dale:11,dale:12} have also used SPH simulations to study the effect of ionizing radiation from the forming massive stars. AMR codes have also been widely used in this context, accounting for a lot of physics and initial and boundary conditions. We note the work of \cite{girichidis:11,girichidis:12a,girichidis:12b} using FLASH who studied the importance of isolated initial conditions in isothermal  cloud core collapse (without stellar feedback).   \cite{federrath:14} also used FLASH to perform simulations of isolated cores with protostellar feedback (jets). \cite{krumholz:07,krumholz:12,hansen:12,myers:13,myers:14,cunningham:11,li:18,offner:17} presented 3D collapse models using ORION with radiative transfer and/or ideal MHD, as well as protostellar feedback (luminosity, outflows). RAMSES has also been extensively used in this approach: \cite{hennebelle:11,commercon:11a} for 100~\msun~ isolated core collapse models with (R)MHD, \cite{lee:18b,lee:18,lee:19} focused on the peak of the IMF using (M)HD 1000~\msun~ isolated core collapse models, \cite{geen:15,geen:16,gavagnin:17} studied the effect of ionizing radiation using the M1 moment method for radiative transfer. 
Last, we note the work of \cite{wang:10} using ENZO with isolated boundary conditions, ideal MHD and protostellar jets. This non-exhaustive list of works done using this approach demonstrates the importance and the utility of such models. 
 These simulations, however, rarely follow the formation of the second Larson core. 
They still rely on the sink particle technique, and on sub-grid models as well, when it comes to modeling proto-stellar jet launching, energy budget at accretion shocks and their associated radiation. 

The last category of simulations deals with isolated collapsing low-mass dense cores, 
 with the goal of following the formation of protostellar disks and ultimately the entire second
collapse without relying on any sub-grid model. This ambitious strategy has already been discussed at length in the previous sections, 
so we will not describe it here again. We also refer the reader to the review by \cite{wurster:18b} on the formation of protostellar disks.

\begin{figure}[h!]
	\begin{center}
		\includegraphics[width=17cm]{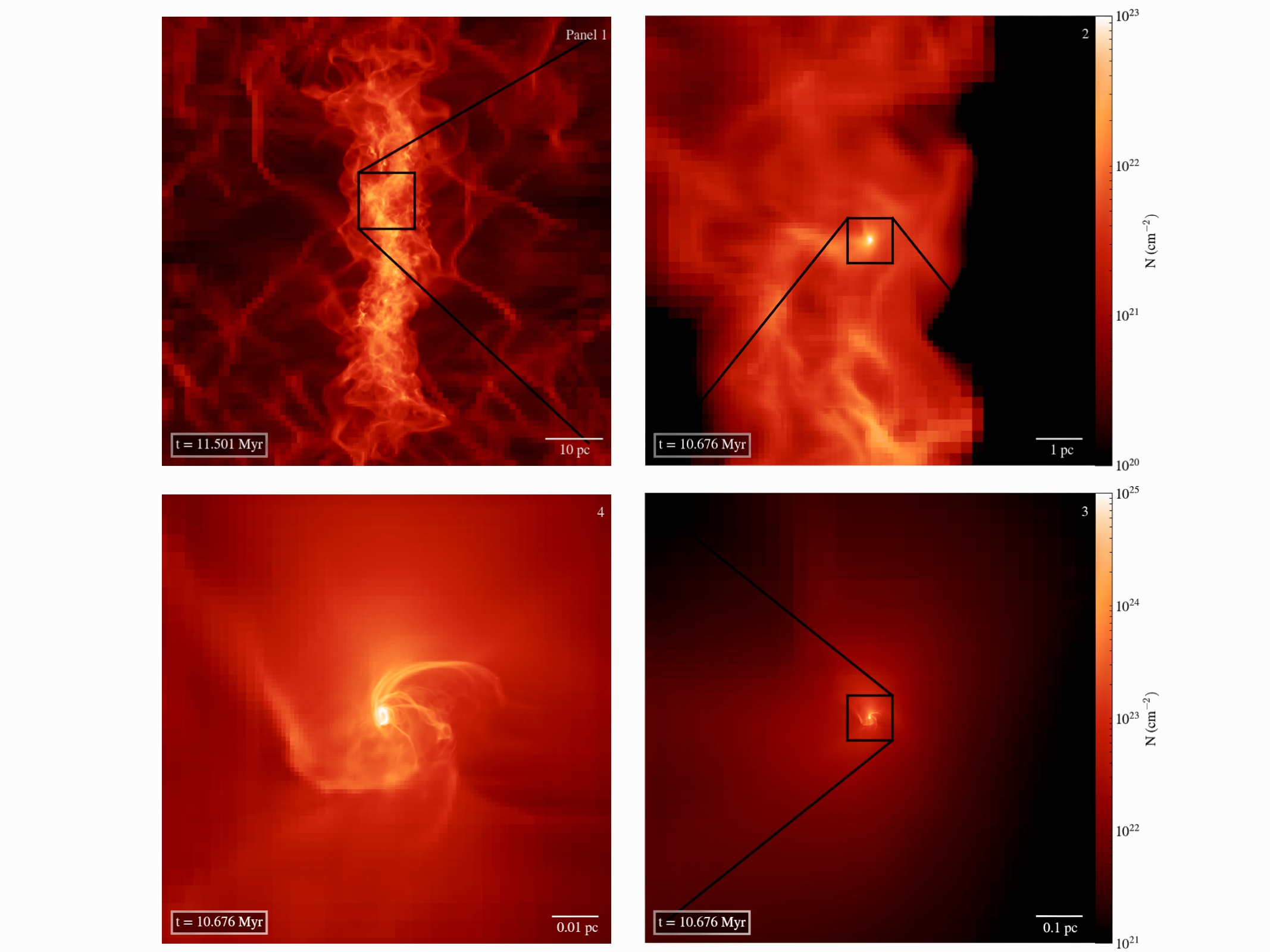}
	\end{center}
	\caption{Simulation of protostellar disk formation within molecular clouds. The column density maps show the entire molecular (upper left panel) and successive zoom within a star-forming dense core.  Figure adapted from \cite{zhang:18} with permission from the authors. \label{fig:zhang}}
\end{figure}

An interesting intermediate strategy has been developed recently with the zoom-in technique. This method is used now routinely in galaxy formation simulations,
for which only a single dark matter halo can be re-simulated at much higher resolution, keeping the entire cosmological environment alive but at a coarser resolution.
Here, the idea is to model the turbulent molecular cloud on large scales, say with a box of size 100 parsecs, to identify a single collapsing core, and to 
zoom-in on the core with much higher resolution and follow the first and possibly the second collapse and the formation of the proto-planetary disk. For instance, \cite{kuffmeier:17,kuffmeier:18} presented RAMSES AMR zoom-in calculations  from an outer scale of 40 pc down to cell sizes of 2 au to study the effect of the environment on the formation and the evolution of protoplanetary disks.  \cite{zhang:18} performed high resolution calculations of the formation and evolution of a star-forming core, obtained by running larger scale calculations of molecular cloud formation \citep{zamora:18}. They used FLASH with a maximum resolution of 25 AU and were able to cover scales from 256 pc to 25 AU (see Fig.~\ref{fig:zhang}). 
\cite{mocz:17} performed moving mesh AREPO calculations from a box size of about 5 pc down to a maximum resolution of $\simeq 4$ AU ($\simeq 5\times10^{-5}$~pc) to study the magnetic fields morphology from  large to dense core scales. At larger scale, \cite{padoan:17} performed  a simulation of supernova-driven turbulence using RAMSES, with a box size of 250 pc and a maximum resolution of $0.0076$~pc. At kpc scales, \cite{hennebelle:18} (FRIGG project with RAMSES,  simulation of the formation of self-gravitating cores) and \cite{seifried:17} (SLICC-Zoom project with FLASH, simulations of the formation of  molecular clouds) presented zoom calculations of stratified Galactic disks down to resolution of $10^{-2}-10^{-3}$ pc. Thanks to the steady increase in CPU power, we expect to see more and more work using the zoom-in technique in the coming years.

These different strategies are all in different ways very ambitious and address different problems in the theory of star formation. 
 In addition, a compromise needs to be found between physical realism and computational efficiency.
On one hand, we need to include magnetic fields, radiation fields and complex chemistry, but on the other hand,
we need many small resolution elements and many time steps. Moreover, in the context of high performance computing,
modern supercomputers are very hard to use at full efficiency when complex grid geometries is used in conjunction with expensive and demanding algorithms.
There is a clear tendency for simulation projects deployed on the largest supercomputers in the world to use simple grid geometry, 
like Cartesian meshes and periodic boxes, with a highly simplified physical model. These large scale simulations are however very interesting to explore
statistical aspects and large inertial range for turbulent flows.
 For example, \cite{federrath:13} performed $4096^3$ hydrodynamic isothermal turbulence simulations using FLASH with periodic boundary condition for more than 40 000 time-steps on  32  768 CPU  cores. In the framework of the magnetorotational turbulence, \cite{fromang:10,ryan:17} performed high resolution simulations of isothermal shearing boxes using, respectively, ZEUS and a GPU version of RAMSES. 
  
 
In conclusion, all simulations of star formation published in the literature, many of which have been discussed here, explore various corners of
the numerical parameter space: resolution versus box size, statistics versus internal structure, physical realism versus computational speed. 
It is interesting that at the smallest scales of interest here, a ``first principle'' approach is in principle possible and pursued by several groups.
At larger scales, however, star formation simulations share the same kind of limitations as galaxy formation simulations, or climate models and weather forecasting 
simulations, namely a strong dependence of the results on small, unresolved scales. Although bigger computers with more efficient, higher order codes and 
more realistic models will certainly help shed light of the mysteries of star formation, a robust methodology to implement sub-grid models 
and couple them properly to the fluid equations still needs to be invented in our field. Various attempts have been proposed in the context of unresolved
turbulence and star formation from analytical works \citep{krumholz:05b,hennebelle:11b,padoan:11,federrath:12} but the methodology, 
 in the context of the full spectrum of required physical processes, is still at its infancy. 

\section*{Acknowledgments}

We would like to warmly thank our two referees, Richard Klein and Christoph Federrath, for their help
in reviewing the manuscript. We thank Matthew Bate, Florent Renaud and Shangjia Zhang for providing adapted figures from their work. 



\bibliographystyle{frontiersinHLTH&FPHY} 

\bibliography{biblio,romain}

\end{document}